\documentclass{article}

\usepackage{microtype}
\usepackage{graphicx}
\usepackage{subcaption}
\usepackage{booktabs} 
\usepackage{algorithm}
\usepackage{algpseudocode}
\usepackage{url}
\usepackage{braket}
\usepackage{xcolor}
\usepackage{multirow} 
\usepackage{array} 
\usepackage{tikz}
\usepackage{tabularray}
\usepackage{enumitem}
\usepackage{amssymb}
\usepackage{hyperref}
\usetikzlibrary{positioning,arrows.meta,fit,calc,shapes.misc}

\usepackage[preprint]{icml2026}

\usepackage{amsmath}
\usepackage{amssymb}
\usepackage{mathtools}
\usepackage{amsthm}

\usepackage[capitalize,noabbrev]{cleveref}

\theoremstyle{plain}

\theoremstyle{definition}

\theoremstyle{remark}

\usepackage[textsize=tiny]{todonotes}

\begin{document}

\twocolumn[
  \icmltitle{Hybrid Action Reinforcement Learning for Quantum Architecture Search}



  \icmlsetsymbol{equal}{*}
  \begin{icmlauthorlist}
    \icmlauthor{Jiayang Niu}{rmit}
    \icmlauthor{Yan Wang}{rmit}
    \icmlauthor{Jie Li}{rmit,sportsbet}
    \icmlauthor{Ke Deng}{rmit}
    \icmlauthor{Azadeh Alavi}{rmit}
    \icmlauthor{Muhammad Usman}{rmit,csiro}
    \icmlauthor{Yongli Ren}{rmit}
  \end{icmlauthorlist}

  \icmlaffiliation{rmit}{School of Computing Technologies, RMIT University, Australia}
  \icmlaffiliation{csiro}{Data61, CSIRO, Australia}
  \icmlaffiliation{sportsbet}{Sportsbet, Australia}

  \icmlcorrespondingauthor{JiayangNiu}{S4068570@student.rmit.edu.au}

  \icmlkeywords{Machine Learning, ICML}

  \vskip 0.3in
]



\printAffiliationsAndNotice{}  

\begin{abstract}
Reinforcement learning--based Quantum Architecture Search (QAS) offers a promising avenue for automating the design of variational quantum circuits, but existing methods typically decouple discrete structure search from continuous parameter optimization, resulting in inefficient or brittle solutions. We propose \textbf{HyRLQAS} (\textbf{H}ybrid-\textbf{A}ction \textbf{R}einforcement \textbf{L}earning for \textbf{Q}uantum \textbf{A}rchitecture \textbf{S}earch), a unified Reinforcement Learning (RL) framework that jointly learns gate placement and parameter initialization within a hybrid discrete--continuous action space, while enabling dynamic refinement of previously placed gates. Trained in a variational quantum eigensolver (VQE) setting, the agent constructs circuits that directly optimize molecular ground-state energies. Across multiple molecular benchmarks, HyRLQAS demonstrates strong and competitive performance against state-of-the-art QAS methods, achieving lower energy errors with fewer gates. Notably, 
HyRLQAS reaches chemical-accuracy–level convergence down to $10^{-8}$ energy error after classical optimization, and this policy-guided initialization reduces the iteration count of downstream classical optimizers. These results demonstrate that hybrid-action reinforcement learning provides a principled and effective mechanism for coupling circuit topology design with optimization-aware parameterization.
\end{abstract}

\section{Introduction}

In the Noisy Intermediate-Scale Quantum (NISQ) era~\cite{bharti2022noisy}, current quantum hardware remains constrained by limited effective circuit depth and significant noise. Consequently, \textit{variational quantum algorithms} (a prominent class of hybrid quantum--classical methods) have emerged as the dominant paradigm for achieving near-term quantum advantage~\cite{mcclean2016theory}. These methods rely on iterative feedback loops between a quantum processor and a classical optimizer, and their effectiveness largely depends on the design of the underlying \textit{Parameterized Quantum Circuit} (PQC)~(or \textit{ansatz}~\cite{mcclean2016theory}). Prominent examples include the Variational Quantum Eigensolver (VQE)~\cite{peruzzo2014variational}, the Quantum Approximate Optimization Algorithm (QAOA)~\cite{farhi2014quantum}, and various variational quantum machine learning models~\cite{schuld2021effect, meyer2023exploiting}. 


Because the performance of variational quantum algorithms critically depends on the choice of 
parameterized quantum circuits (PQCs), considerable effort has been devoted to designing effective 
ansatz structures. While several PQC families have been proposed, such as hardware-efficient~\cite{kandala2017hardware}, chemically inspired~\cite{zeng2023quantum}, and problem-specific~\cite{peruzzo2014variational, skolik2023equivariant} templates, these designs are typically constructed manually or rely on fixed heuristic patterns. However, such hand-crafted circuits often struggle to generalize across different problem instances or system sizes, or to adapt to hardware noise and connectivity constraints, thereby limiting scalability, expressivity, and robustness on realistic NISQ devices~\cite{mcclean2018barren, du2020expressive, jose2022error}. These challenges motivate \textit{Quantum Architecture Search} (QAS)~\cite{zhang2022differentiable}, which aims to automatically design PQC ansatz tailored to a given task and hardware constraints.



While evolutionary methods~\cite{williams1998automated, potovcek2018multi, chivilikhin2020mog}, differentiable approaches~\cite{wu2023quantumdarts} and sampling-based strategies~\cite{tang2021qubit, he2024training, wang2023automated} had been explored in QAS, recent works have focused on reinforcement learning (RL) as a flexible framework for QAS~\cite{ostaszewski2021reinforcement, patel2024curriculum, dutta2025qas, ikhtiarudin2025benchrl} due to its effectiveness, which is also the focus of this paper. 
In RL-based QAS, an agent sequentially constructs a quantum circuit, 
while circuit parameters are optimized by an external routine, and the resulting task-driven performance (e.g., the achieved VQE energy, a QAOA objective value, or a variational quantum learning loss) is used as the reward for policy updates.
Despite promising results, existing RL-based frameworks face three issues: \textbf{i}) the gate placement and continuous parameter optimization are decoupled, overlooking their interaction, including the effect of parameter initialization; 
in most existing RL-based QAS frameworks, the external optimizer is invoked independently in each episode, preventing the reuse of parameter-related experience and often resulting in inefficient optimization with excessive iterations;
and \textbf{iii}) performance remains constrained by the shortcomings of the external optimizer, including sensitivity to initialization (See Appendix~\ref{preliminary_experiment}) and trainability issues at larger circuit depths~\cite{mcclean2018barren, larocca2025barren}.

To overcome these issues, we propose the \textbf{H}ybrid-\textbf{A}ction \textbf{R}einforcement \textbf{L}earning for \textbf{Q}uantum \textbf{A}rchitecture \textbf{S}earch (\textbf{HyRLQAS}) framework. HyRLQAS expands the agent’s action space to jointly learn \emph{where} to place gates (discrete actions) and \emph{how} to initialize their parameters (continuous actions) at circuit construction time, thereby capturing the interaction between circuit topology and parameter initialization (issue \textbf{i}). We also explored end-to-end parameter learning but found it unstable in practice (Sec.~\ref{sec:ablationA}), motivating our focus on parameter initialization. Rather than replacing variational optimization, HyRLQAS learns informed policy-guided initialization distributions that reuse optimization knowledge across episodes 
and reduce the number of iterations required by downstream classical optimizers (issue~\textbf{ii}), and improve post-optimization performance
(issue~\textbf{iii}). Furthermore, we analyze this effect using the Quantum Neural Tangent Kernel (QNTK) and its dynamic variant (dQNTK)~\cite{liu2022representation}, which characterize optimization dynamics in variational quantum circuits. Our analysis shows that policy-guided initialization improves kernel conditioning and gradient propagation, providing a principled explanation for the observed convergence and robustness gains. 

The contributions of this work are as follows:
\begin{itemize}
    \item The \textbf{HyRLQAS} that jointly searches circuit structures via discrete gate placement and parameter initialization through continuous actions under a unified policy. Our implementation is available at \url{https://anonymous.4open.science/r/HyRLQAS-4ED1}.
    \item A policy-guided mechanism that provides informed parameter initializations for external variational optimizers, together with a QNTK- and dQNTK-based analysis that explains the resulting improvements in optimization stability and performance.
    \item A comprehensive evaluation of HyRLQAS through extensive experiments on 
    standard molecular VQE benchmarks, 
    demonstrating improvements over baseline methods and attributing performance gains to both policy-guided initialization and hybrid-action–induced architectural improvements.
\end{itemize}


\section{Related Work}


Reinforcement learning~\cite{mcclean2018barren, van2016deep} has been increasingly explored for quantum architecture search (QAS), where task-specific reward signals are derived from variational quantum algorithms (VQAs) such as VQE, state preparation, or combinatorial optimization~\cite{khairy2020learning}.  In the context of VQE, \citet{ostaszewski2021reinforcement} pioneered the use of curriculum learning in RL-based ansatz search, enabling agents to progressively construct circuits under an energy-driven reward. Building on this paradigm, \citet{patel2024curriculum} proposed CRLQAS, which enhances curriculum learning with structured state encodings, dynamic masking of illegal actions, and an improved external optimizer to boost sample efficiency and robustness. More recently, \citet{kundu2025tensorrl} introduced TensorRL-QAS, which integrates tensor-network techniques into RL-based QAS by employing matrix product state (MPS) approximations to warm-start circuit construction, significantly improving scalability to larger qubit systems. In addition, RL has also been applied to QAS with alternative objectives such as fidelity or task-specific costs, enabling applications in quantum classification and optimization~\cite{dutta2025qas, ikhtiarudin2025benchrl}. 

In parallel, other research improves scalability by structuring the search space with reusable subcircuits (gadgets or blocks). For example, block-based ansatz construction for combinatorial optimization~\cite{turati2025automated}, gadget reinforcement learning with composite gates~\cite{kundu2024reinforcement}, and systematic discovery of Clifford gadgets to accelerate RL agents~\cite{olle2025scaling}. In addition, non-RL frameworks have been developed to reduce computational cost, such as the hardware-tested QAS of~\citet{du2022quantum} and the training-free QAS of \citet{he2024training}, which rely on proxy-based evaluations instead of circuit training.  Beyond architecture search, reinforcement learning has also been employed for circuit-level optimization. For instance, Quarl~\cite{li2024quarl} formulates circuit optimization as a sequential decision process and leverages graph neural networks with a structured action space to guide local optimization decisions.

Overall, existing methods predominantly focus on \emph{circuit architecture search}, while largely overlooking the role of parameter decision-making during circuit construction.
The few works that address parameter-related aspects, such as \citet{peng2025breaking}, employ reinforcement learning for parameter initialization under \emph{fixed circuit structures}, and therefore do not consider joint optimization of architecture and parameters.
This separation motivates a learning paradigm in which structural choices and continuous parameter decisions are made simultaneously.

From a methodological perspective, the reinforcement learning community has extensively studied \emph{hybrid action spaces}, which model discrete decisions and continuous parameters jointly.
Representative works include parameterized action spaces~\cite{hausknecht2015deep}, hybrid actor--critic architectures~\cite{fan2019hybrid}, and Hybrid Action Representation (HyAR)~\cite{lihyar}, which learns compact representations of structured discrete--continuous actions.
These advances provide a natural methodological foundation for formulating quantum architecture search as a hybrid-action problem.
Motivated by this perspective, we propose \textbf{HyRLQAS}, which casts quantum architecture search as a unified hybrid-action reinforcement learning task that jointly optimizes circuit structure and parameter initialization.



\section{Methodology}
\begin{figure*}[t]
    \centering
    \includegraphics[width=0.8\linewidth]{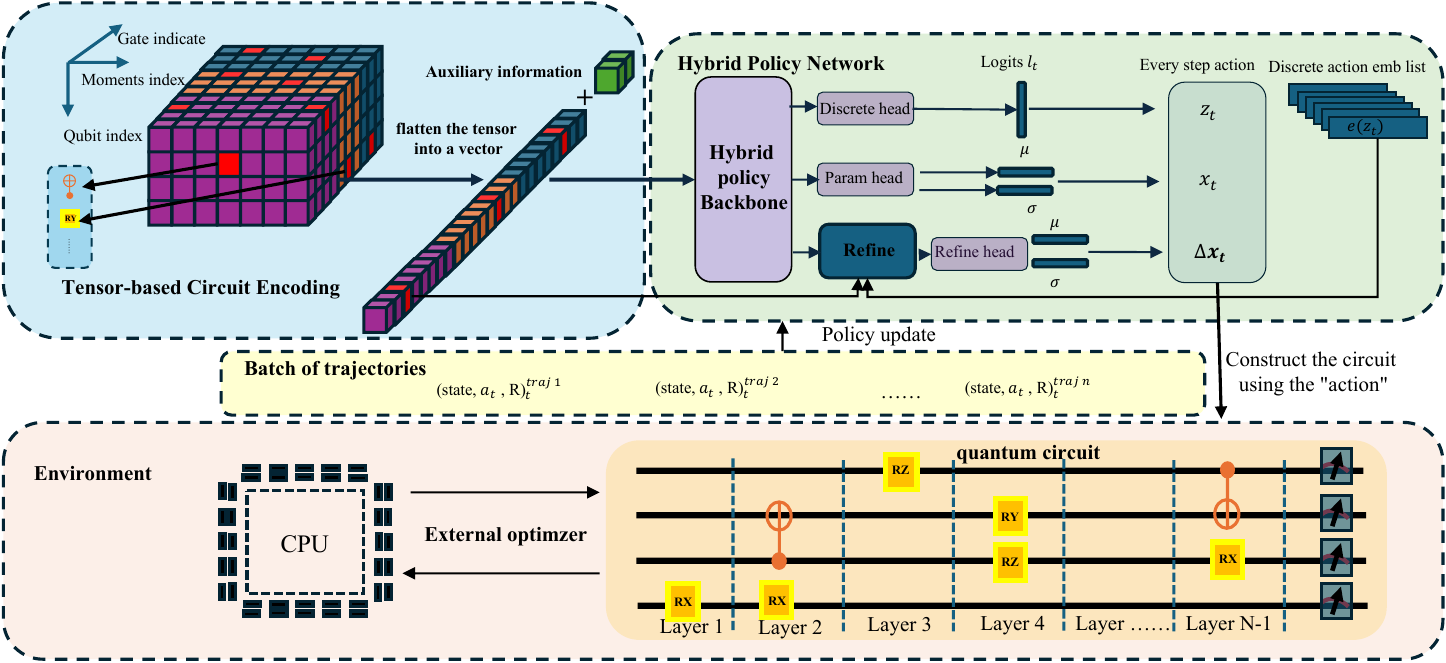}
    \caption{Overview of the proposed \textbf{HyRLQAS} framework. A tensor-based circuit encoding summarizes the partially constructed circuit, which is fed into a hybrid policy network to generate discrete gate choices, continuous parameter initialization, and refinement actions. The environment executes the resulting circuit and provides feedback, and collected trajectories are used to update the policy.}
    \label{fig:overall_HyRLQAS}
\end{figure*}

\subsection{HyRLQAS Framework}
The proposed reinforcement learning-based solution framework, Hybrid-Action Reinforcement Learning for Quantum Architecture Search (HyRLQAS) (Figure~\ref{fig:overall_HyRLQAS}). Consistent with~\cite{ostaszewski2021reinforcement, patel2024curriculum}, HyRLQAS formulates the problem as a reinforcement learning process. It employs an agent that learns a policy $\pi(a_t|s_t)$ to select an action $a_t$ given a state $s_t$. Here, $s_t$ denotes the environment's state, represented by an intermediate circuit from the Tensor-based Circuit Encoding module (see Appendix~\ref{appendix:TBCE}).

Each episode begins with an empty circuit $c_0$, which is encoded as the initial state $s_0$. At each decision step $t$, the agent observes the current state $s_t$ and, guided by its current policy $\pi(a_t|s_t)$, selects an action $a_t$ to incrementally extend the circuit. An action $a_t$ consists of a discrete action $z_t$ and a continuous parameter $x_t$. Specifically, $z_t$ specifies the type and location of the gate to be inserted at step $t$, while $x_t$ defines a continuous parameter for the initial rotation angle of the newly introduced parameterized gate. Upon applying $a_t$, the reward $r(s_t, a_t)$ is observed, and the state transitions to $s_{t+1}$. This interaction continues until a termination condition is satisfied. Through repeated interactions, the agent learns a policy that maps circuit states to actions $(z_t, x_t)$, with the objective of constructing quantum circuits that maximize the expected cumulative reward.

At each step, the quality of the currently constructed circuit is evaluated using a Variational Quantum Eigensolver (VQE) procedure. Once a candidate circuit has been specified up to step $t$, an external classical optimizer is invoked to refine all variational parameters associated with the circuit so as to minimize the target energy. Given the optimized parameters $\vec{\theta}$, the circuit prepares a quantum state $|\psi(\vec{\theta})\rangle$, and the energy of the circuit at step $t$ is evaluated as the expectation value of the molecular Hamiltonian $H$:
\begin{equation}
\label{eq:1}
E_t = \langle \psi(\vec{\theta}) \mid H \mid \psi(\vec{\theta}) \rangle .
\end{equation}

The energy $E_t$ is computed after each circuit extension step and subsequently transformed into a reward signal $r_t$ for reinforcement learning. 
It reflects how close the variational state prepared by the intermediate circuit is to the ground-state energy of the target molecule.
(See Appendix~\ref{appendix:energy} for details on the Hamiltonian representation and the expectation value computation.)

Based on the per-step energy evaluation described above, we use a reward function $R$~\cite{ostaszewski2021reinforcement} based on energy feedback:
\begin{equation}
\label{eq:reward_fn}
r(s_t, a_t) =
\begin{cases}
+5, & \text{if } E_t < \xi, \\
-5, & \text{if } t \geq \ell  \ \text{and}\ E_t \geq \xi, \\
\max\left( \dfrac{E_{t-1} - E_t}{E_{t-1} - E_{\min}}, -1 \right), & \text{otherwise}.
\end{cases}
\end{equation}

The threshold $\xi$ is dynamically adjusted through feedback-driven curriculum learning. Episode termination occurs under one of two conditions:
(i) the energy of the constructed circuit falls below the threshold $\xi$, or
(ii) the episode reaches its maximum length $\ell$, which is determined by a random halting mechanism and is upper-bounded by a predefined limit $L$, without satisfying the energy criterion.

Accordingly, the first two cases in Eq.~\ref{eq:reward_fn} correspond to terminal rewards and are assigned only at episode completion.
Specifically, a positive terminal reward is given when the target energy threshold is achieved, whereas a negative terminal reward is assigned if the episode terminates without meeting the threshold.
The third case provides stepwise reward shaping during the episode, delivering feedback proportional to the relative energy improvement achieved at each decision step.
Detailed descriptions of the feedback-driven curriculum learning strategy, the random halting mechanism are provided in Appendix~\ref{appendix:RH}, Appendix~\ref{appendix:FDCL}.

\subsection{Hybrid Policy Optimization}

An action $a$ comprises a discrete component $z$ and a continuous parameter $x$. This formulation aligns with the framework of Parameterized Action Markov Decision Processes (PAMDPs) \cite{masson2016reinforcement}, which are defined by a continuous state space $S \subseteq \mathbb{R}^{n}$ and parameterized action spaces. In a PAMDP, a discrete action is chosen from a finite set, each associated with its own continuous parameters. Although $x$ can generally be regarded as a parameter of $z$, we adopt a two-stage action selection process: first selecting a parameterized action, and then selecting its corresponding continuous parameters. In our setting, the choices of $z$ (e.g., the type and placement of a gate) and $x$ (e.g., the initial rotation angle of the gate) are decoupled. We thus define a stochastic policy that maps a state $s$ to both the discrete action $z$ and the continuous parameter $x$ as follows:

$$
\pi(z,x \mid s) = \pi^d (z \mid s) \, \pi^a (x \mid s),
$$

where $\pi^d(z \mid s)$ denotes the discrete-action policy and $\pi^a(x \mid s)$ the continuous-parameter policy. Their parameters are $\phi^d$ and $\phi^a$, respectively.

The discrete action space includes three types of single--qubit rotation gates ($R_X$, $R_Y$, $R_Z$) acting on any qubit, as well as two--qubit CNOT gates acting on ordered qubit pairs.
For a circuit with $N$ qubits, this results in a discrete action space of size $3N + 2\binom{N}{2}$. To ensure physical validity and avoid redundant constructions, the environment dynamically identifies illegal operations—such as consecutive identical rotation gates or inverted CNOT pairs—based on the current circuit state and masks them out from the discrete action set prior to sampling (see Appendix~\ref{appendix:IA}).

The continuous components of the action are conditionally defined given the discrete gate decision.
Specifically, if the selected discrete action $z_t$ corresponds to a parameterized single--qubit rotation, the initialization parameter $x_t$ is sampled from a Gaussian distribution,
\begin{equation}
x_t \sim \mathcal{N}\!\left(\mu_t, \sigma_t^2\right),
\end{equation}
where the mean $\mu_t$ and standard deviation $\sigma_t$ are produced by the policy network.
No continuous initialization is generated when the selected gate is non-parameterized.

When a new gate is added to a quantum circuit, the parameters of the existing rotation gates must be updated to preserve optimal performance. This refinement is modeled by adding a Gaussian-distributed increment to each current parameter vector $\boldsymbol{x}_t$. The update is given by $\boldsymbol{x}_t \leftarrow \boldsymbol{x}_t + \eta \Delta\boldsymbol{x}_t$, where $\eta$ is a step-size parameter and the increment $\Delta\boldsymbol{x}_t$ is drawn from a reference distribution $\pi_{\mathrm{ref}}(\cdot \mid s_t, z_t, {x}_t)$.
Only entries corresponding to valid parameterized gates are retained, with all others masked out.
This conditional factorization enables coherent parameter refinement aligned with the current circuit structure and the newly inserted gate, supporting structured exploration in the continuous parameter space.

The hybrid policy is optimized using a policy-gradient method based on REINFORCE, and the overall training and interaction procedure is summarized in Algorithm~\ref{alg:hyrlqas} in Appendix~\ref{HASID}.

\section{Experiments}

Following \cite{ostaszewski2021reinforcement,patel2024curriculum,kundu2025tensorrl}, the proposed HyRLQAS is evaluated on quantum architecture search (QAS) under the variational quantum eigensolver (VQE) setting, using molecular ground-state energy estimation as the downstream task. 
Experiments are conducted on standard molecular VQE benchmarks (\textsc{LiH}, \textsc{BeH$_2$}, and \textsc{H$_2$O}) in the STO-3G basis with symmetry-reduced qubit encodings following prior work~\citep{patel2024curriculum}. 
All main results are obtained under noiseless simulation to assess the intrinsic optimization capability of each method, while supplementary experiments under realistic noise models are reported in the appendix~\ref{appendix:Exper_noisy}; detailed molecular configurations are provided in Appendix~\ref{appendix:MoleConf}.


We compare against representative QAS baselines spanning three categories: \textbf{i) reinforcement learning--based} TensorRL(CRLQAS)~\citep{kundu2025tensorrl}, CRLQAS~\citep{patel2024curriculum}, VanillaRL~\citep{ostaszewski2021reinforcement}, and BenchRL-QAS (DQN-rank and TPPO)~\citep{ikhtiarudin2025benchrl}; \textbf{ii) differentiable} quantumDARTS~\citep{wu2023quantumdarts}; and \textbf{iii) sampling-based} TF-QAS~\cite{he2024training}. For fair comparison under TensorRL's warm-start protocol, we also evaluate our method within the TensorRL framework (TensorRL(HyRLQAS)). All methods are evaluated on all benchmark systems, except TF-QAS, which is included only on BeH$_2$--6 following prior work\footnote{For TF-QAS and quantumDARTS, we report results from the original publications due to the lack of public implementations.}.

For each learning-based method, the checkpoint achieving the lowest training energy is selected and evaluated under noiseless simulation, reporting ground-state energy error, circuit depth, and gate counts. Detailed hardware and hyperparameter settings are provided in Appendix~\ref{appendix:Exp_setting}. 

\begin{table*}[ht]
\centering
\tiny
\caption{Comparison of QAS methods under noiseless simulations on benchmark molecular systems. We report the ground-state energy error, circuit depth, and total gate count for several baselines and our proposed HyRLQAS.}
\label{Tab:Main_Exp}
\resizebox{0.9\textwidth}{!}{
\begin{tblr}{
  cells = {c},
  cell{2}{1} = {r=8}{},
  cell{2}{7} = {r=8}{},
  cell{10}{1} = {r=8}{},
  cell{10}{7} = {r=8}{},
  cell{2}{3} = {bg=red!90},
  cell{3}{3} = {bg=red!40},
  cell{7}{5} = {bg=red!40},
  cell{3}{4} = {bg=red!90},
  cell{3}{5,6} = {bg=red!40},
  cell{4}{4-6} = {bg=red!90},
  cell{10}{3} = {bg=red!40},
  cell{17}{3} = {bg=red!90},
  cell{11}{4-6} = {bg=red!90},
  cell{12}{4,5} = {bg=red!40},
  cell{12}{6} = {bg=red!90},
  cell{6,10}{9} = {bg=red!90},
  cell{2,17}{9} = {bg=red!40},
  cell{11}{10-12} = {bg=red!90},
  cell{3}{10,12} = {bg=red!90},
  cell{3}{11} = {bg=red!40},
  cell{4}{10,12} = {bg=red!40},
  cell{4}{11} = {bg=red!90},
  cell{12}{10-12} = {bg=red!40},
  hline{1,2,10,18} = {-}{0.12em},
  vline{1,2,3,7,8,9,13} = {-}{0.12em},
}
Molecule & Method           & Error          & Depth & CNOT & ROT & Molecule & Method           & Error        & Depth & CNOT & ROT            \\
\textsc{LiH-4}    & Ours HyRLQAS     & 1.2×10\^{}-8   & 19    & 11   & 22  & \textsc{BEH$_2$-6} & Ours HyRLQAS & 6.3×10\^{}-8 & 10    & 14   & 12             \\
         & TensorRL(HyRLQAS)& 5.1×10\^{}-6   & 5     & 4    & 9  &           & TensorRL(HyRLQAS)& 8.8×10\^{}-6 & 5     & 7    & 5              \\
         & TensorRL(CRLQAS) & 6.5×10\^{}-4   & 5     & 2    & 8  &           & TensorRL(CRLQAS) & 8.4×10\^{}-5 & 6     & 5    & 8              \\
         & CRLQAS           & 9.6×10\^{}-5   & 17    & 11   & 19  &          & CRLQAS           & 7.3×10\^{}-8 & 21    & 26   & 8              \\
         & Vanilla RL       & 3.5×10\^{}-4   & 11    & 8    & 11  &          & Vanilla RL       & 2.3×10\^{}-8 & 37    & 32   & 23             \\
         & DQN-rank         & 1.5×10\^{}-3   & 15    & 4    & 23  &          & DQN-rank         & 9.8×10\^{}-7 & 24    & 16   & 28             \\
         & TPPO             & 1.2×10\^{}-3   & 24    & 14   & 26  &          & TPPO             & 2.7×10\^{}-7 & 23    & 22   & 19             \\
         & quantumDARTS     & 1.7×10\^{}-4   & 34    & 34   & 50  &          & TF-QAS           & 1.8×10\^{}-3 & N/A   & N/A  & 57(total gate) \\
\textsc{LiH-6}    & Ours HyRLQAS     & 5.9×10\^{}-4   & 23    & 27   & 24  & \textsc{H$_2$O-8}  & Ours HyRLQAS     & 1.7×10\^{}-4 & 23    & 25   & 14             \\
         & TensorRL(HyRLQAS)& 8.1×10\^{}-4   & 5     & 5    & 4   &          & TensorRL(HyRLQAS)& 5.1×10\^{}-4 & 4     & 6    & 4              \\
         & TensorRL(CRLQAS) & 1.3×10\^{}-3   & 6     & 6    & 4   &          & TensorRL(CRLQAS) & 9.1×10\^{}-4 & 11    & 8    & 10             \\
         & CRLQAS           & 1.5×10\^{}-3   & 29    & 28   & 22  &          & CRLQAS           & 1.1×10\^{}-3 & 30    & 25   & 42             \\
         & Vanilla RL       & 3.7×10\^{}-3   & 36    & 38   & 25  &          & Vanilla RL       & 1.1×10\^{}-3 & 59    & 67   & 65             \\
         & DQN-rank         & 1.2×10\^{}-3   & 24    & 19   & 43  &          & DQN-rank         & 9.2×10\^{}-4 & 35    & 34   & 30             \\
         & TPPO             & 1.0×10\^{}-3   & 25    & 23   & 42  &          & TPPO             & 9.5×10\^{}-4 & 56    & 76   & 29             \\
         & quantumDARTS     & 2.9×10\^{}-4   & 54    & 52   & 80  &          & quantumDARTS     & 3.1×10\^{}-4 & 64    & 68   & 151     
\end{tblr}
}
\end{table*}

\subsection{Comparision to Baselines}

As summarized in Table~\ref{Tab:Main_Exp}, \textbf{HyRLQAS} achieves a substantial performance improvement on \textsc{LiH--4}, reducing the energy error from the $10^{-5}$--$10^{-4}$ range attained by existing QAS methods to the $10^{-8}$ level, without increasing circuit depth or gate count.
Such an improvement is rarely observed in prior RL-based QAS studies and cannot be attributed to simply using deeper or more expressive circuits. Beyond this small-system regime, HyRLQAS consistently maintains an advantage over competing RL-QAS baselines on \textsc{LiH--6}, \textsc{BeH$_2$--6}, and \textsc{H$_2$O--8}. Across these larger molecules, HyRLQAS attains lower energy errors while using comparable or fewer gates, indicating that its performance gains persist as system size and circuit depth increase, rather than overfitting to shallow circuits or small instances.

When integrated into the TensorRL framework, \textbf{TensorRL (HyRLQAS-based)} further improves energy accuracy over \textbf{TensorRL (CRLQAS-based)} while consistently producing the most compact circuits among all TensorRL variants, typically achieving the shortest depths and lowest gate counts. Overall, these results show that hybrid-action reinforcement learning enables HyRLQAS to achieve a superior accuracy--complexity tradeoff across molecular systems of increasing size.

\subsection{Ablation Study}
\subsubsection{A: Effect of Each Component}
\label{sec:ablationA}
This ablation study evaluates the contribution of the three core components in \textbf{HyRLQAS}: 
(i) the \emph{hybrid action space}, 
(ii) the \emph{refinement head}, and 
(iii) the \emph{external optimizer}. 

We construct three reduced variants by disabling one component at a time: \textbf{A1: w/o Hybrid Space (Discrete-only):} removes continuous parameter learning and optimizes only gate placement. \textbf{A2: w/o Refine Head:} disables retroactive parameter updates, keeping previously assigned parameters fixed. \textbf{A3:w/o External Optimizer:} evaluates agent-generated parameters directly without post-hoc global optimization. All variants are trained under identical settings and evaluated on \textsc{LiH--4} and \textsc{LiH--6}. We report the final energy estimation error together with circuit depth, gate count, and parameter count (Table~\ref{tab:ablation_a}).

\begin{table}[ht]
\centering
\caption{Ablation Study A: performance comparison of HyRLQAS and its reduced variants on \textsc{LiH-4} and \textsc{LiH-6} molecules.}
\tiny
\label{tab:ablation_a}
\resizebox{\linewidth}{!}{
\begin{tblr}{
  cells = {c},
  cell{1}{1} = {r=2}{},
  cell{1}{2} = {c=4}{},
  cell{1}{6} = {c=4}{},
  hline{1,3,7} = {-}{0.15em},
  hline{2} = {2-10}{0.15em},
  vline{1,2,6,10} = {-}{0.15em},
}
Method                 & \textsc{LiH-4}            &        &       &       & \textsc{LiH-6}             &          &       &       \\
                       & Error                     & Depth  & CNOT  & ROT   & Error                      & Depth    & CNOT  & ROT   \\
A1                     & \underline{2.1×10\^{}-6}  & 24     & 18    & 19    & \underline{1.6×10\^{}-3}   & 31       & 20    & 29    \\
A2                     & 2.9×10\^{}-5              & 25     & 24    & 13    & 2.2×10\^{}-3               & 41       & 37    & 33    \\
A3                     & 3.8×10\^{}-2              & 26     & 17    & 23    & 7.3×10\^{}-2               & 29       & 19    & 36    \\
FULL                   & \textbf{1.2×10\^{}-8}     & 19     & 11    & 22    & \textbf{5.9×10\^{}-4}      & 23       & 27    & 24   
\end{tblr}
}
\end{table}

\paragraph{Results and Analysis:}
The full HyRLQAS consistently achieves the lowest energy error with balanced circuit complexity on both systems. 
Removing the hybrid action space significantly degrades performance, confirming the importance of learning parameter initialization at circuit construction time jointly with gate placement.
Disabling the refinement head leads to an even larger accuracy drop, highlighting the necessity of maintaining parameter consistency as the circuit grows. 
Finally, removing the external optimizer causes a marked performance degradation, indicating that hybrid-action RL primarily contributes by identifying high-quality initial parameter regions, while the external optimizer performs local fine-tuning for final convergence. Overall, these results demonstrate that HyRLQAS derives its performance gains from the \emph{combination} of hybrid-action learning and refinement-based parameter adaptation, with the external optimizer providing the final stage of precision optimization.

\subsubsection{B: Circuit Structure vs. Parameter Initialization}
We further disentangle the performance gains of \textbf{HyRLQAS} by isolating the effects of 
(i) circuit structure, 
(ii) parameter initialization, and 
(iii) refinement-based parameter adjustment.
To decouple structural and parametric effects, we freeze the trained policy from the main experiment and evaluate multiple controlled variants under identical circuit structures (B1--B4). Any performance difference therefore reflects only initialization or refinement effects rather than architectural variation.

We consider five variants summarized in Table~\ref{tab:ablation_b}:
\textbf{B1:} full HyRLQAS with policy-guided initialization and refinement;
\textbf{B2:} learned initialization without refinement accumulation;
\textbf{B3:} zero initialization;
\textbf{B4:} random initialization;
and \textbf{B5:} a discrete-only baseline without continuous actions.
Specifically, B2 isolates the effect of refinement-based parameter adjustment by retaining policy-guided parameter initialization for newly added gates, while disabling retroactive updates to parameters assigned in previous steps. 
All variants are evaluated using COBYLA on \textsc{LiH-4} and \textsc{LiH-6}, reporting the best energy estimation error.
To quantify the benefit of hybrid-action learning, we additionally report the \emph{relative error reduction} of each variant with respect to the discrete-only baseline B5.

\begin{table}[ht]
\centering
\caption{Ablation Study B: Best energy estimation errors on \textsc{LiH-4} and \textsc{LiH-6}. ER vs.\ B5 (\%) indicates the percentage error reduction relative to the discrete-only baseline B5.}
\label{tab:ablation_b}
\tiny
\resizebox{\linewidth}{!}{
\begin{tblr}{
  cell{1}{1} = {c=5},
  cells = {c},
  hline{1,2,3,8} = {-}{0.15em},
  vline{1,6} = {-}{0.15em},
  vline{2} = {2-8}{0.15em},
}   
Results and Analysis &                                     &                  &                  &              \\
Experiment ID        & Setting Description                 & \textsc{LiH-4} Error      & \textsc{LiH-6} Error      & ER vs B5 (\%) \\
B1                   & HyRLQAS                             & 1.26×10\^{}-8    & 5.91×10\^{}-4    & 99.4 / 63.7  \\
B2                   & HyRLQAS w/o refinement accumulation & 5.69×10\^{}-7    & 1.02×10\^{}-3    & 73/37.4      \\
B3                   & HyRLQAS with Zero init              & 3.31×10\^{}-7    & 8.18×10\^{}-4    & 84.3 / 49.8  \\
B4                   & HyRLQAS with Random init            & 1.07×10\^{}-7    & 7.82×10\^{}-4    & 94.9/52      \\
B5                   & HyRLQAS w/o Hybrid space            & 2.11×10\^{}-6    & 1.63×10\^{}-3    & * / *        
\end{tblr}
}
\end{table}

\paragraph{Results Analysis}
As shown in Table~\ref{tab:ablation_b}, hybrid-action learning improves performance through both circuit structure and parameter initialization.
Even without learned initialization, the hybrid-action model (B3) significantly outperforms the discrete-only baseline (B5), achieving error reductions of 84.3\% on \textsc{LiH-4} and 49.8\% on \textsc{LiH-6}. This demonstrates that hybrid-action learning already provides a strong architectural advantage by identifying more effective circuit structures, as further illustrated by the qualitative circuit visualizations in Appendix~\ref{appendix:final_arch_A1}.
Comparing B1 with B3 and B4 further highlights the role of learned initialization. While zero and random initialization (B3 and B4) yield similar performance, policy-guided initialization (B1) consistently achieves lower errors, indicating that the gains arise from learned initialization rather than random parameter choices. Finally, disabling refinement accumulation (B2) results in pronounced performance degradation, with errors exceeding those of zero or random initialization, confirming that refinement-based parameter adjustment is essential for effectively exploiting the continuous action branch. 
To further understand the role of refinement, we analyze its parameter dynamics during circuit construction in Appendix~\ref{app:refine_magnitude}.

\begin{figure*}[t]
    \centering
    \begin{subfigure}{0.45\linewidth}
        \centering
        \includegraphics[width=\linewidth]{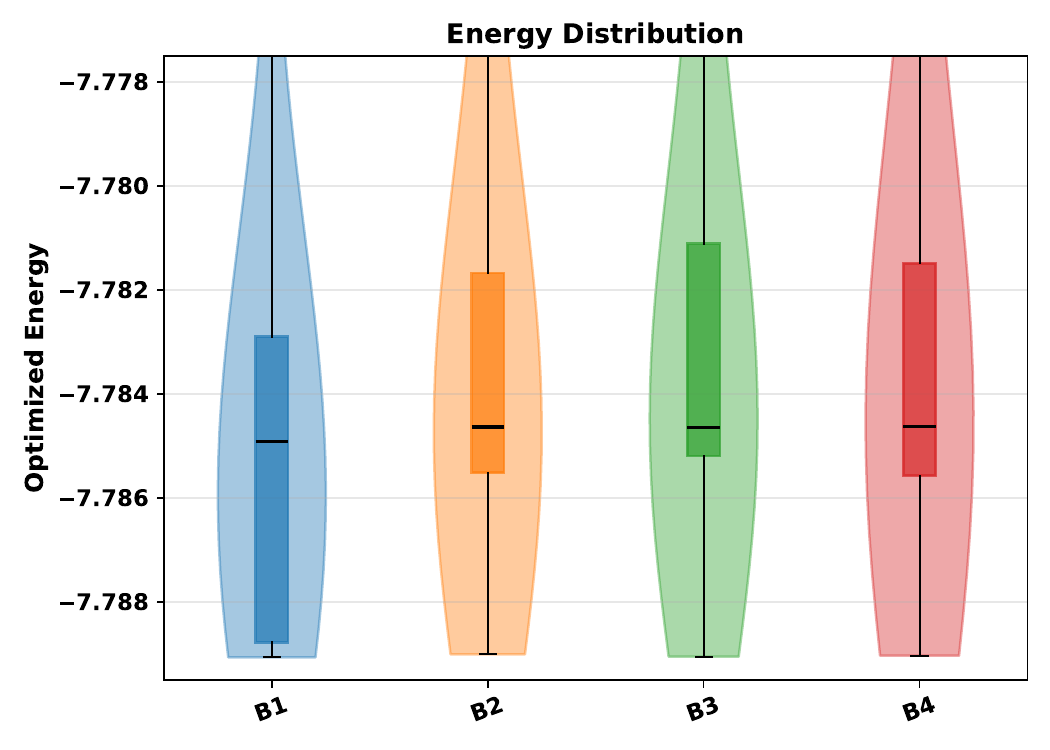}
    \end{subfigure}
    \begin{subfigure}{0.45\linewidth}
        \centering
        \includegraphics[width=\linewidth]{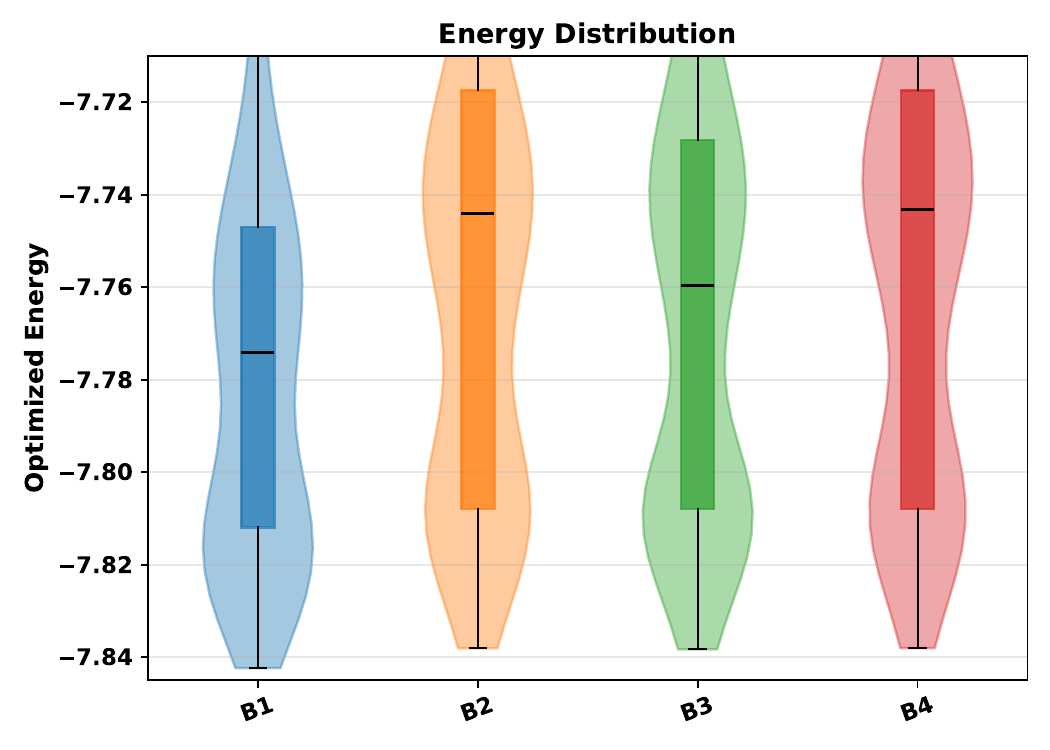}
    \end{subfigure}
    \caption{Final energy distributions of COBYLA-optimized circuits under four hybrid-action variants, evaluated using the same frozen policy to ensure identical circuit structures. \textsc{LiH-4} (left) and \textsc{LiH-6} (right).
    }
    \label{fig:energy_4Strategy}
\end{figure*}

\paragraph{Visualization and Statistical Validation}
Figure~\ref{fig:energy_4Strategy} compares the final energy distributions of B1--B4 on \textsc{LiH-4} and \textsc{LiH-6} using the same frozen hybrid-action policy, ensuring identical circuit structures across settings. The full model (B1) consistently concentrates probability mass toward lower-energy regions, whereas removing refinement accumulation (B2) induces a clear rightward shift, indicating degraded solution quality. 
To quantify this difference, we perform two-sample Kolmogorov--Smirnov (KS) tests. As an illustrative case, comparing B1 with the zero-initialized variant (B3) yields $D=0.133$ ($p=4.0\times10^{-8}$) on \textsc{LiH-4} and $D=0.224$ ($p=2.2\times10^{-22}$) on \textsc{LiH-6}, with similar trends observed across other settings.
These results confirm that policy-guided initialization, especially when coupled with refinement, induces a statistically distinct optimization landscape with a higher likelihood of reaching low-energy solutions than zero initialization. 

In addition to quantitative metrics, we visualize the final optimized circuit architectures produced by HyRLQAS and CRLQAS. Qualitative comparisons of the optimized structures are provided in Appendix~\ref{appendix:final_arch}, illustrating systematic differences in gate allocation and circuit motifs.




\begin{figure*}[ht]
\centering

\begin{subfigure}{\textwidth}
\centering
\begin{subfigure}{0.32\textwidth}
    \includegraphics[width=\linewidth]{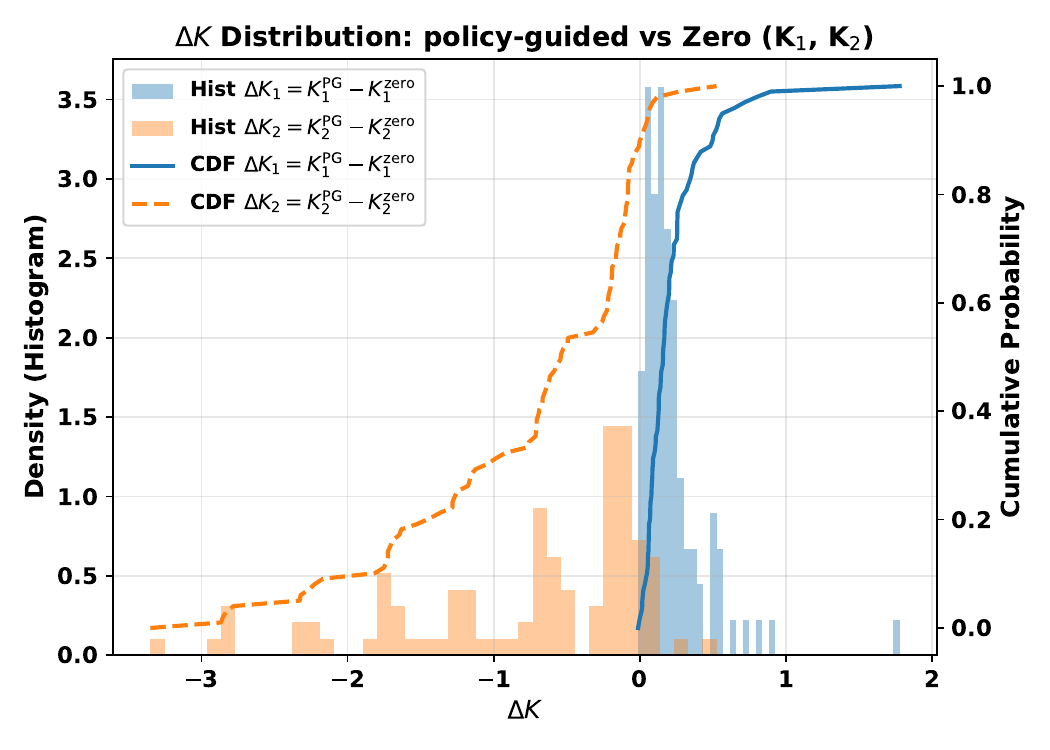}
    \caption{Policy-Guided vs Zero}
    \label{fig:deltaK_pg_zero_LiH4}
\end{subfigure}
\begin{subfigure}{0.32\textwidth}
    \includegraphics[width=\linewidth]{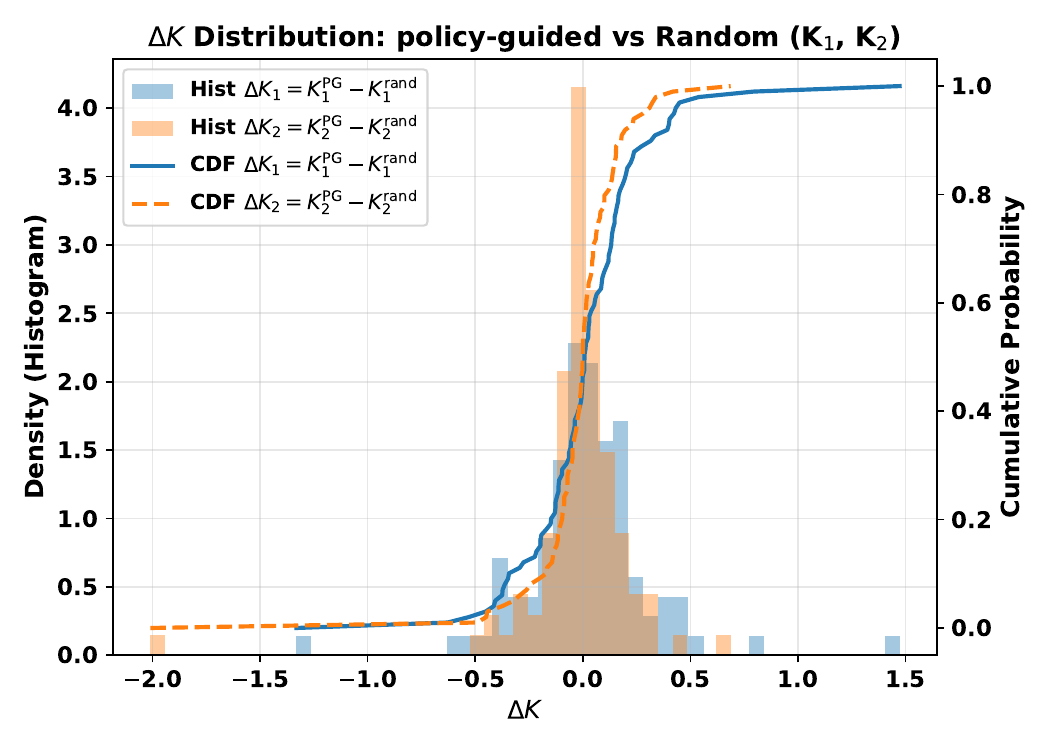}
    \caption{Policy-Guided vs Random}
    \label{fig:deltaK_pg_random_LiH4}
\end{subfigure}
\begin{subfigure}{0.32\textwidth}
    \includegraphics[width=\linewidth]{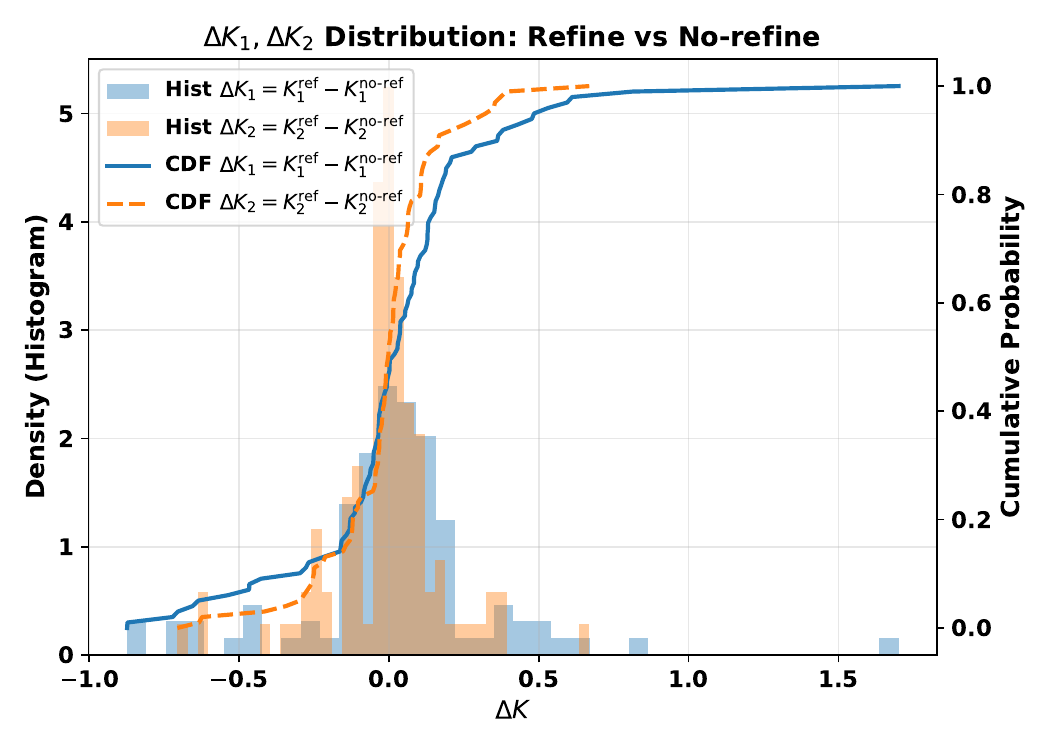}
    \caption{Refine vs No-refine}
    \label{fig:deltaK_ref_noref_LiH4}
\end{subfigure}
\caption{$\Delta K_1$ and $\Delta K_2$ on identical \textsc{LiH-4} circuits, comparing Policy-Guided initialization against Zero, Random, and no-refine variants.}
\label{fig:block_LiH4}
\end{subfigure}

\vspace{1.0ex} 

\begin{subfigure}{\textwidth}
\centering
\begin{subfigure}{0.32\textwidth}
    \includegraphics[width=\linewidth]{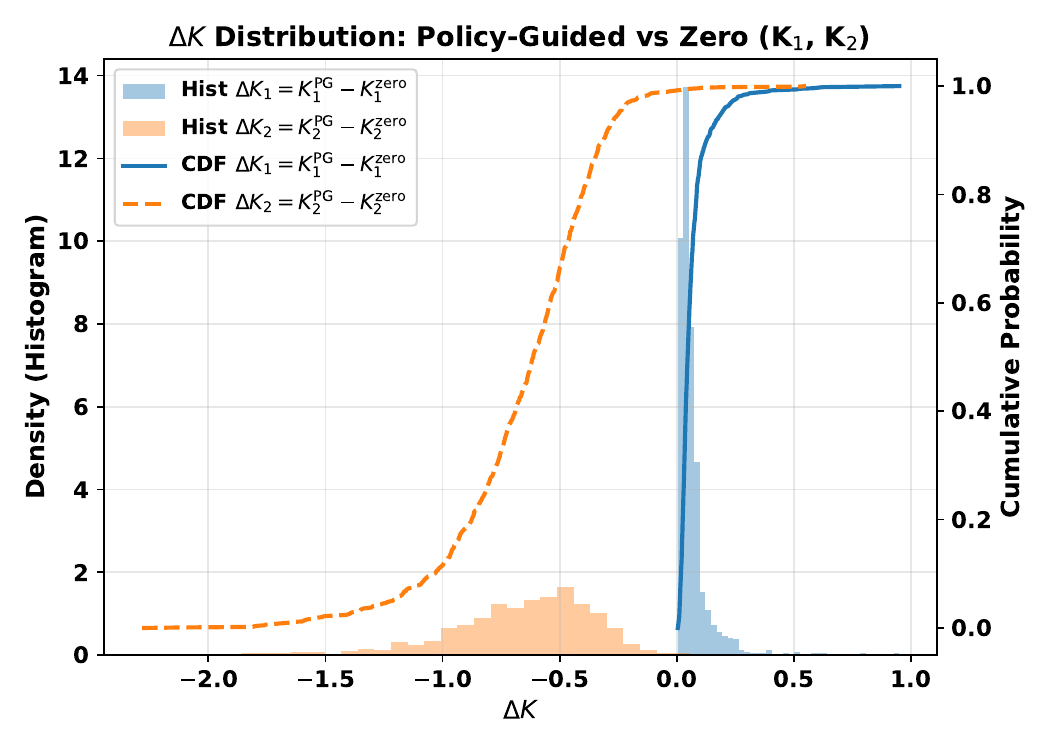}
    \caption{Policy-Guided vs Zero}
    \label{fig:deltaK_pg_zero_LiH6}
\end{subfigure}
\begin{subfigure}{0.32\textwidth}
    \includegraphics[width=\linewidth]{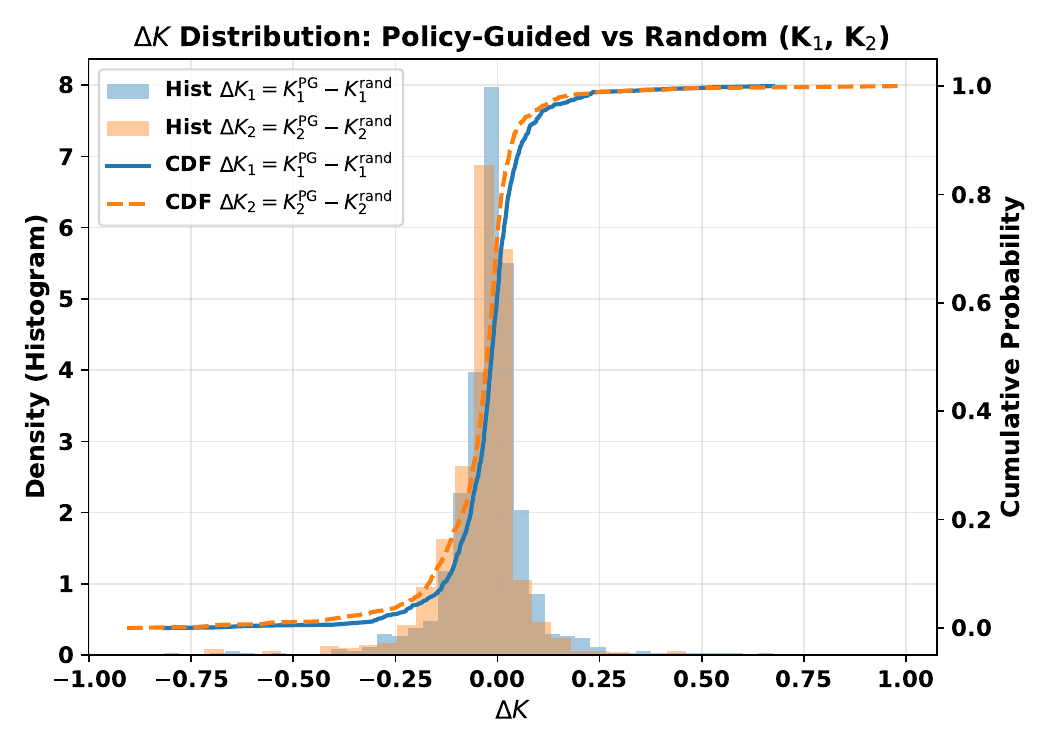}
    \caption{Policy-Guided vs Random}
    \label{fig:deltaK_pg_random_LiH6}
\end{subfigure}
\begin{subfigure}{0.32\textwidth}
    \includegraphics[width=\linewidth]{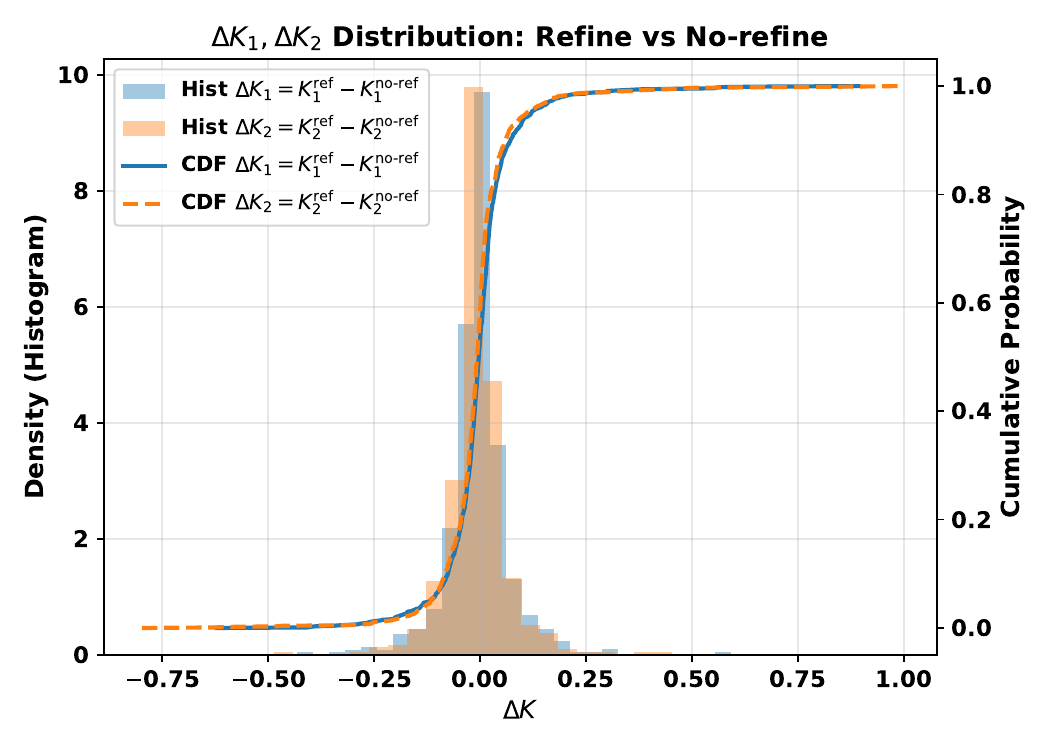}
    \caption{Refine vs No-refine}
    \label{fig:deltaK_ref_noref_LiH6}
\end{subfigure}
\caption{$\Delta K_1$ and $\Delta K_2$ on identical \textsc{LiH-6} circuits, comparing Policy-Guided initialization against Zero, Random, and no-refine variants.}
\label{fig:block_LiH6}
\end{subfigure}

\vspace{1.0ex}

\begin{subfigure}{\textwidth}
\centering
\begin{subfigure}{0.23\textwidth}
    \includegraphics[width=\linewidth]{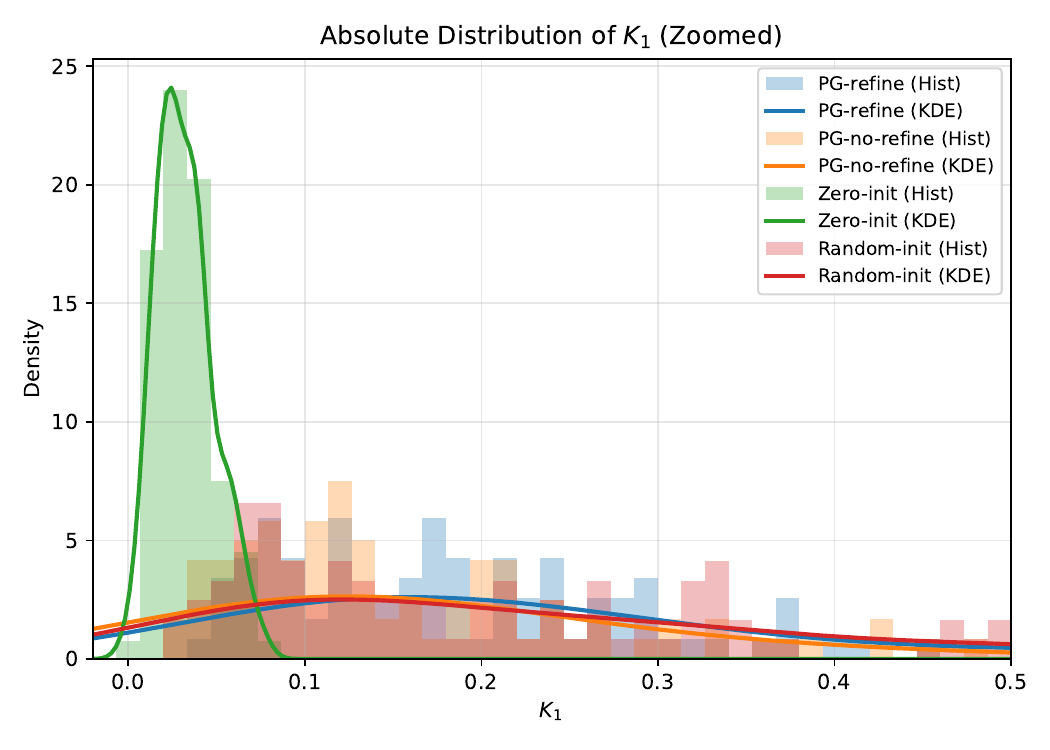}
    \caption{$K_1$ (\textsc{LiH-4})}
    \label{fig:K1_abs_LiH4}
\end{subfigure}
\begin{subfigure}{0.23\textwidth}
    \includegraphics[width=\linewidth]{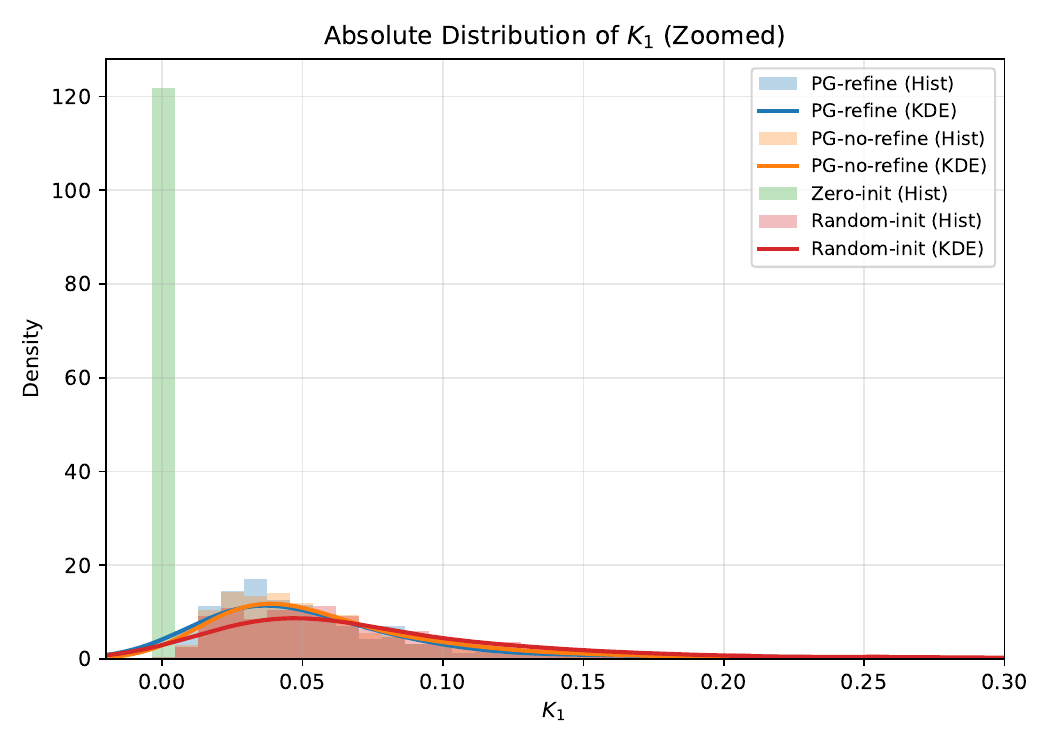}
    \caption{$K_1$ (\textsc{LiH-6})}
    \label{fig:K1_abs_LiH6}
\end{subfigure}
\begin{subfigure}{0.23\textwidth}
    \includegraphics[width=\linewidth]{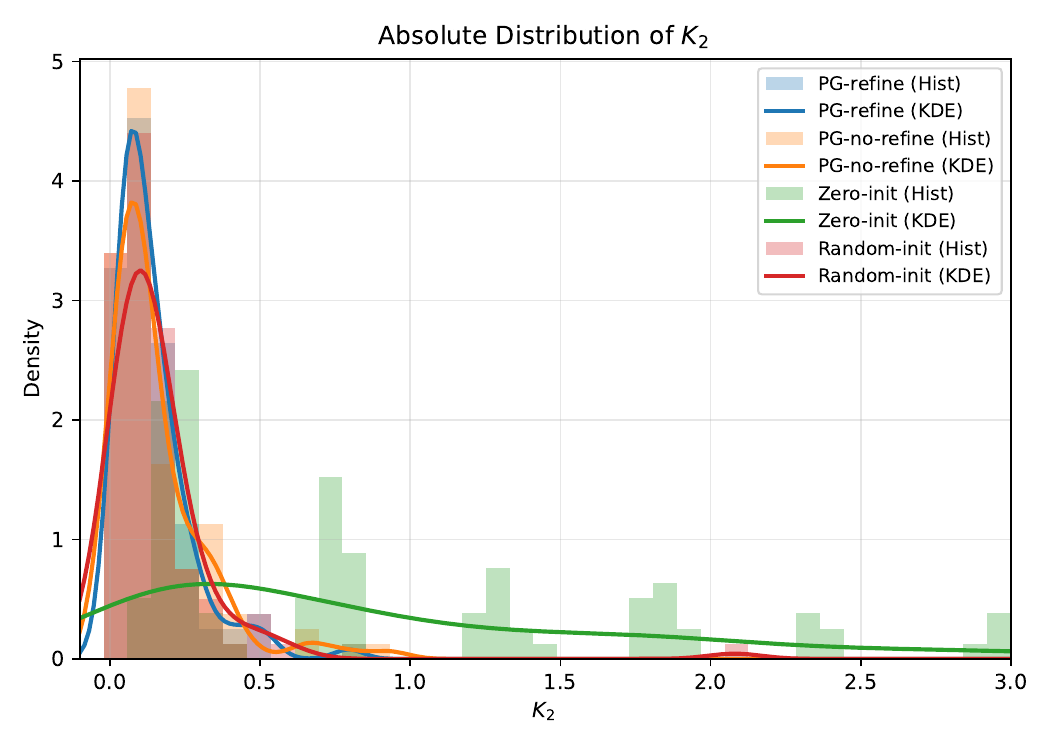}
    \caption{$K_2$ (\textsc{LiH-4})}
    \label{fig:K2_abs_LiH4}
\end{subfigure}
\begin{subfigure}{0.23\textwidth}
    \includegraphics[width=\linewidth]{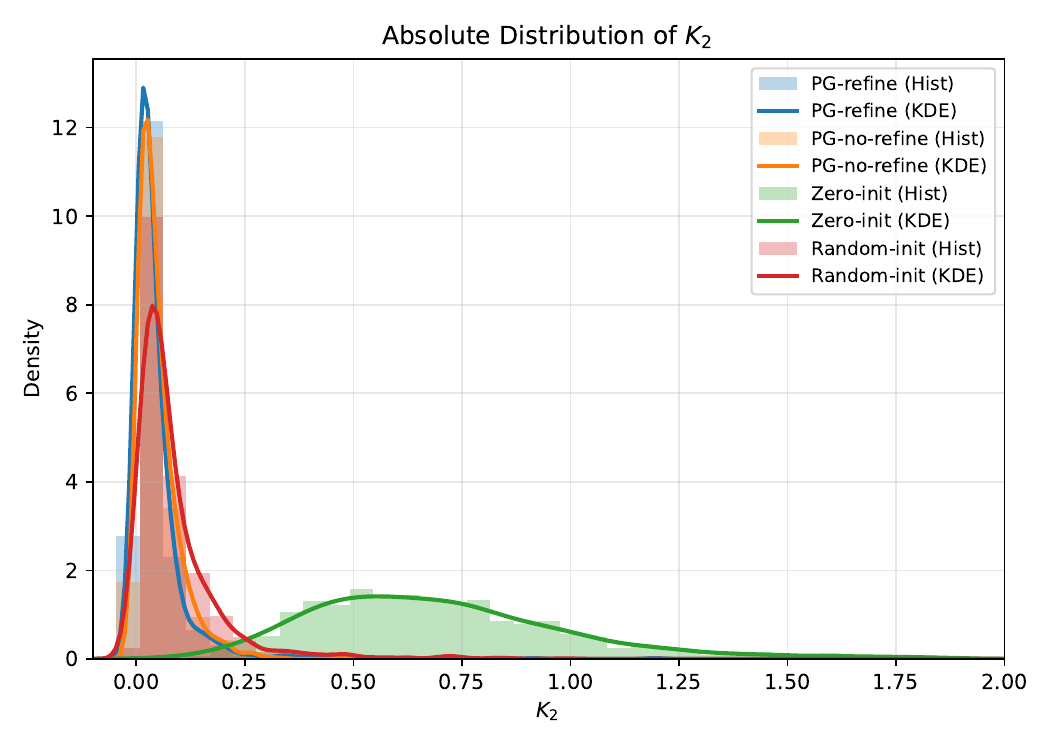}
    \caption{$K_2$ (\textsc{LiH-6})}
    \label{fig:K2_abs_LiH6}
\end{subfigure}
\caption{distributions of raw $K_1$ and $K_2$ values for all variants, illustrating the absolute kernel scale and concentration.}
\label{fig:block_absK}
\end{subfigure}

\caption{
QNTK analysis for \textsc{HyRLQAS} on \textsc{LiH-4} and \textsc{LiH-6},  aggregated over \textbf{1000 circuit architectures} generated by a frozen policy. }
\label{fig:qntk_all}
\end{figure*}

\subsection{Trainability Analysis}
To probe \emph{why} policy-guided parameter decisions improve downstream optimization, we analyze the optimization geometry of HyRLQAS-generated circuits through the \textbf{Quantum Neural Tangent Kernel (QNTK)}~\cite{liu2022representation}. We compute the first- and second-order QNTK statistics $(K_1, K_2)$ prior to any classical parameter optimization (Formal definitions are provided in Appendix~\ref{app:qntk}), and compare how different parameterization strategies reshape the local trainability landscape.

\paragraph{Setup.}
We evaluate circuits generated by a \emph{frozen} HyRLQAS policy and aggregate results over \textbf{1000} independently constructed architectures for each molecule. To isolate optimization effects from architectural bias, all comparisons are performed on \emph{identical circuit structures} and differ only in parameterization: (i) \textbf{Policy-Guided (PG)} initialization with refinement, (ii) \textbf{PG no-refine} Policy-Guided initialization without refinement, (iii) \textbf{zero} initialization, and (iv) \textbf{random} initialization. For each circuit, we report both the pairwise kernel differences $\Delta K_1,\Delta K_2$ (Figs.~\ref{fig:block_LiH4}--\ref{fig:block_LiH6}) and the absolute distributions of $K_1$ and $K_2$ (Fig.~\ref{fig:block_absK}).

\paragraph{Findings A.} \emph{PG vs.\ zero initialization shows a clear separation in kernel geometry.} On both \textsc{LiH-4} and \textsc{LiH-6}, $\Delta K_1$ is predominantly positive while $\Delta K_2$ is predominantly negative for PG compared to zero (Figs.~\ref{fig:deltaK_pg_zero_LiH4},~\ref{fig:deltaK_pg_zero_LiH6}), indicating that PG places circuits in a regime with stronger first-order sensitivity and reduced second-order curvature effects. This trend is consistent with the absolute $K_1$ distributions, where zero initialization concentrates more mass near low-$K_1$ values
(Figs.~\ref{fig:K1_abs_LiH4},~\ref{fig:K1_abs_LiH6}), suggesting degraded local trainability under identity-centered parameterization.

\paragraph{Findings B.} \emph{PG is less distinguishable from random or no-refine under $\Delta K$ alone, motivating inspection of absolute distributions.}  In contrast, the $\Delta K_1$ and $\Delta K_2$ comparisons between PG and random, as well as between refine and no-refine, show substantial overlap (Figs.~\ref{fig:deltaK_pg_random_LiH4}--\ref{fig:deltaK_ref_noref_LiH4} and \ref{fig:deltaK_pg_random_LiH6}--\ref{fig:deltaK_ref_noref_LiH6}), implying that these variants occupy similar first-order kernel neighborhoods. However, the absolute $K_2$ distributions reveal a more informative and consistent signature. Across both \textsc{LiH-4} and \textsc{LiH-6}, PG produces a \emph{sharper and more concentrated} $K_2$ distribution with a mild positive shift (Figs.~\ref{fig:K2_abs_LiH4},~\ref{fig:K2_abs_LiH6}), while random and no-refine variants exhibit visibly broader spreads and heavier tails. \emph{Moreover, the peak height and width of the $K_2$ distribution follows a consistent ordering, with PG-refine attaining the highest and narrowest peak, followed by PG-no-refine, and then random initialization.} This ordering indicates that refinement progressively increases the fraction of circuits operating in a stable and well-conditioned second-order regime.

Taken together, these observations suggest that while multiple strategies may yield comparable first-order geometry, Policy-Guided refinement systematically regularizes second-order kernel structure, providing a coherent explanation for its consistently superior post-optimization performance observed in our previous experiments.

\subsection{Efficiency Analysis}
\label{Exp_warm_up_time}

\begin{table}[]
\centering
\caption{Wall-clock time per step (mean $\pm$ std, in milliseconds), averaged over 1000 evaluation episodes. The Env.\ step includes Optimizer time.}
\label{tab:timing}
\footnotesize
\begin{tabular}{lccc}
\toprule
System & Inference & Env.\ step & Optimizer \\
\midrule
\textsc{LiH-4} & $2.58 \pm 0.04$ & $86.45 \pm 7.43$ & $79.69 \pm 7.39$ \\
\textsc{LiH-6} & $2.66 \pm 0.05$ & $197.02 \pm 25.11$ & $187.33 \pm 24.64$ \\
\bottomrule
\end{tabular}
\end{table}
\begin{table}[t]
\centering
\caption{Action inference time per step (mean $\pm$ std, in milliseconds), averaged over 1000 evaluation episodes.}
\label{tab:inference}
\footnotesize
\setlength{\tabcolsep}{2pt}
\begin{tabular}{lcccc}
\toprule
RL method & CRLQAS & DQN-Rank & TPPO & HyRLQAS \\
\midrule
\textsc{LiH-4} & $1.3 \pm 0.03$ & $0.92 \pm 0.12$ & $1.35 \pm 0.03$ & $2.6 \pm 0.04$ \\
\textsc{LiH-6} & $1.3 \pm 0.04$ & $0.93 \pm 0.09$ & $1.41 \pm 0.04$ & $2.7 \pm 0.05$ \\
\bottomrule
\end{tabular}
\vspace*{-0.5cm}
\end{table}

We profile the per-step wall-clock cost of \textbf{HyRLQAS} by decomposing each circuit construction step into policy inference, environment-side circuit construction, and external classical optimization (Table~\ref{tab:timing}).
Across both \textsc{LiH-4} and \textsc{LiH-6}, the external optimizer overwhelmingly dominates the runtime, while policy inference incurs a small and nearly constant overhead (about $2.6$ms per step).
As further shown in Table~\ref{tab:inference}, although \textbf{HyRLQAS} has slightly higher inference cost than other RL-based QAS baselines due to its hybrid discrete–continuous action space, this overhead remains negligible relative to optimization and does not constitute a bottleneck.

Motivated by this observation, we evaluate whether \emph{policy-guided initialization} can reduce optimization cost.
Under identical circuit structures, policy-guided initialization reduces the number of COBYLA iterations by 33\% on \textsc{LiH-4} and 46\% on \textsc{LiH-6}, while achieving comparable or lower final energies. These results show that learned initialization substantially eases downstream classical optimization, which is the dominant contributor to runtime. 

\section{Conclusion}
We propose \textbf{HyRLQAS}, a hybrid-action reinforcement learning framework that jointly optimizes quantum circuit structure and continuous parameters during circuit construction. By integrating discrete gate placement with policy-guided parameter initialization and refinement, HyRLQAS outperforms discrete-only QAS methods and yields more effective variational circuits. Ablation studies show that hybrid-action learning improves both architectural search and optimization efficiency by providing informed parameterization that reduces reliance on classical optimizers. QNTK analysis further links these gains to more favorable optimization geometry. Future work will investigate fully end-to-end approaches that eliminate external optimizers.

\bibliography{example_paper}
\bibliographystyle{icml2026}

\newpage
\appendix
\onecolumn
\section{Appendix: Parameter Initialization Sensitivity}
\label{preliminary_experiment}

To verify that the optimized performance is sensitive to different initialization strategies when circuits become deep, we conduct a sensitivity study under the same VQE objective as in the main experiments, where performance is measured by the optimized energy after variational optimization.
To demonstrate this, we repeatedly optimize the same circuit under three initialization strategies and record the mean, variance, maximum, and minimum of the resulting energies. We compare the following initialization strategies~\cite{peng2025breaking}:

\begin{itemize}
    \item \textbf{Near-zero}: all parameters are initialized close to zero with small Gaussian perturbations ($\sigma=10^{-3}$). 
    This setting follows the default initialization used in prior RL-based QAS frameworks, while the small perturbation is introduced to show that even minimal deviations can lead to noticeably different optimized energies.
    
    \item \textbf{Random}: each parameterized gate is initialized by sampling its parameter uniformly from the interval $[-\pi, \pi]$. 
    This common random initialization serves as a reference to quantify how much performance variation different initialization strategies can introduce.
    
    \item \textbf{Near-random}: parameters are first initialized using the random strategy and then perturbed with small Gaussian noise ($\sigma=10^{-3}$). 
    This setting is designed to verify that the sensitivity to slight perturbations in parameter initialization is a general phenomenon, not limited to near-zero configurations.
\end{itemize}

The circuits used in this study are generated by the policy obtained from the \textbf{CRLQAS} framework~\cite{patel2024curriculum} during its training process. 
Specifically, we use the policy checkpoint saved at 30\% of the total training episodes. This selection is not based on performance considerations but rather on the observation that, at this stage, the generated circuits have already reached sufficient depth for analyzing initialization sensitivity.  

We employ the COBYLA optimizer with a maximum of 1000 iterations. For each initialization strategy, 100 independent optimization runs are performed on the same circuit to collect statistical measures. To ensure generality, this experiment is conducted across three molecular Hamiltonians of increasing circuit widths: \textsc{LiH} (4 qubits), \textsc{LiH} (6 qubits), and \textsc{H\textsubscript{2}O} (8 qubits). We present the aggregated results, including the mean, standard deviation, minimum, and maximum of the optimized energies.

\begin{figure}[t]
    \centering
    \begin{subfigure}{0.32\linewidth}
        \centering
        \includegraphics[width=\linewidth]{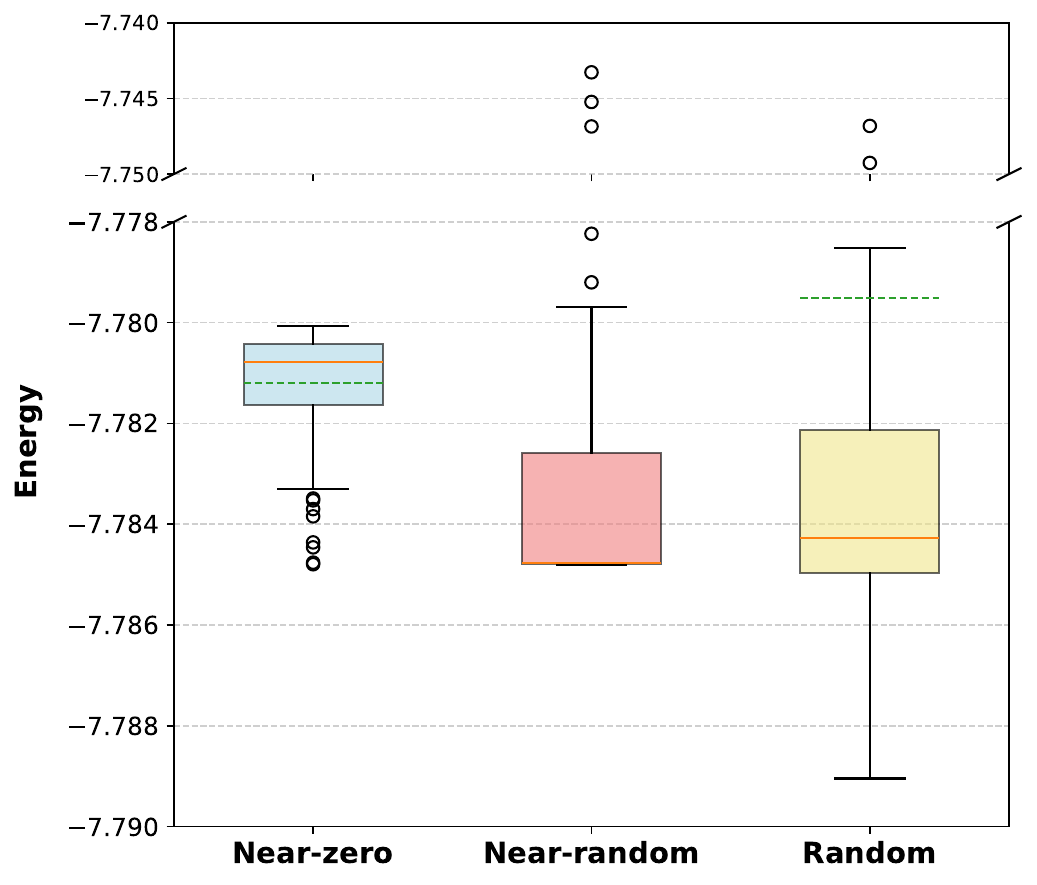}
        \caption{\textsc{LiH} 4-qubit}
        \label{fig:LiH4_appendix}
    \end{subfigure}
    \hfill
    \begin{subfigure}{0.32\linewidth}
        \centering
        \includegraphics[width=\linewidth]{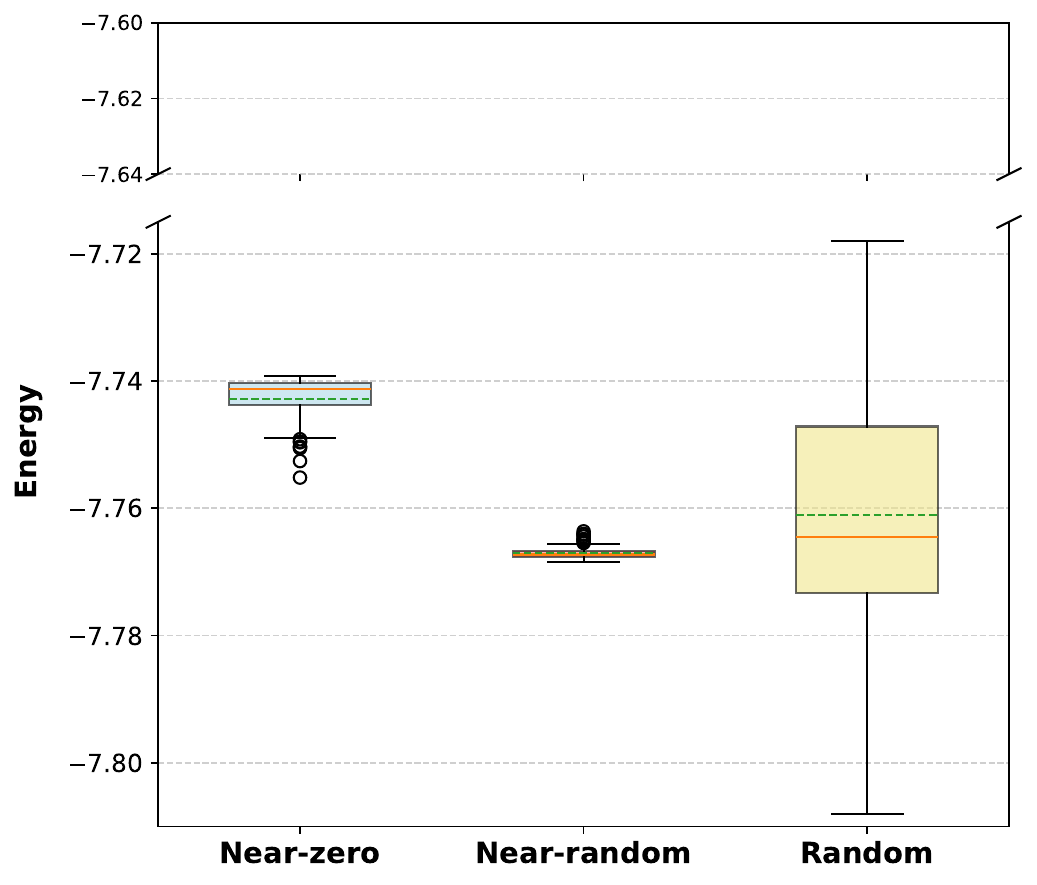}
        \caption{\textsc{LiH} 6-qubit}
        \label{fig:LiH6_appendix}
    \end{subfigure}
    \hfill
    \begin{subfigure}{0.32\linewidth}
        \centering
        \includegraphics[width=\linewidth]{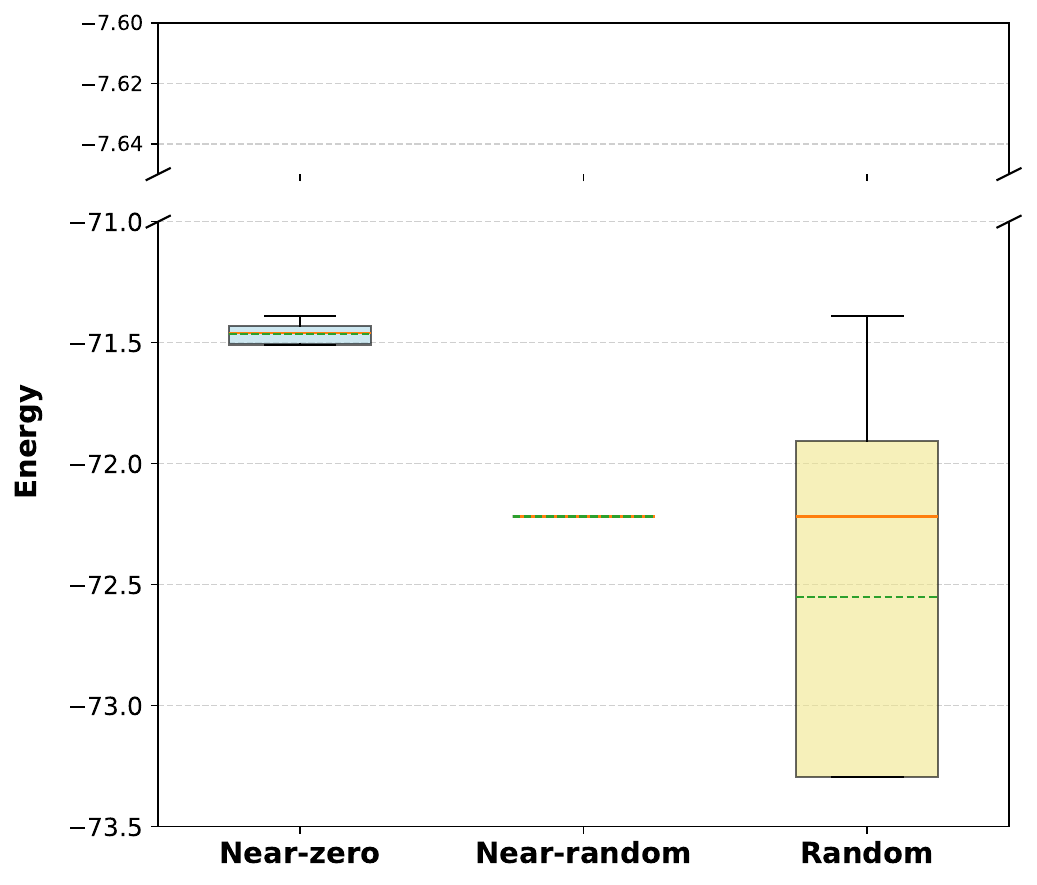}
        \caption{\textsc{H$_2$O} 8-qubit}
        \label{fig:LiH8_appendix}
    \end{subfigure}
    \caption{Distribution of final energies under three initialization strategies for the same circuit. \textit{Near-zero}: parameters are initialized close to zero with small Gaussian noise.}
    \label{fig:init_boxplots_appendix}
\end{figure}

The collected statistics and visual distributions together reveal clear trends in initialization sensitivity. 
As shown in Figure~\ref{fig:init_boxplots_appendix} and Table~\ref{tab:Energy_dist_three_strategies}, different initialization strategies exert a pronounced influence on the optimized energy values, highlighting that circuit optimization is highly sensitive to initialization. 
For example, in the \textsc{LiH-4} system, the final energies obtained under different initialization strategies exhibit clearly separated distributions, indicating that even slight perturbations in the initial parameters can lead to noticeable variations in the optimized energy. 
This demonstrates that parameter initialization plays a critical role in the convergence behavior of variational optimization. 
Although the variance for the \textit{Near-random} setting in the \textsc{H\textsubscript{2}O}-8 case is reported as zero, this does not indicate superior stability; rather, it implies that the optimization stagnated and failed to improve beyond its initial value. 
These observations further underscore the necessity of appropriate initialization strategies for achieving effective energy minimization and stable learning dynamics.

\begin{table}[t]
\centering
\caption{Energy statistics of COBYLA optimization under three initialization strategies (Near-zero, Random, and Near-random) across different molecular Hamiltonians. 
Each entry reports the mean, standard deviation, minimum, and maximum of the optimized energies over 100 independent runs.}
\label{tab:Energy_dist_three_strategies}
\scriptsize
\begin{tblr}{
  cells = {c},
  cell{2}{2} = {r=3}{},
  cell{5}{2} = {r=3}{},
  cell{8}{2} = {r=3}{},
  vline{1,2,3,7} = {-}{0.12em},
  hline{1, 11} = {-}{0.12em},
  hline{2,5,8} = {-}{0.12em}
}
Strategy    & Molecular & energy mean~ & energy std & min energy & max energy \\
Near-zero   & \textsc{LiH-4}     & -7.7812      & 0.0012     & -7.7769    & -7.7848     \\
Random      &           & -7.7795      & 0.0142     & -7.7127    & -7.7890     \\
Near-random &           & -7.7757      & 0.0186     & -7.7260    & -7.7848     \\
Near-zero   & \textsc{LiH-6}     & -7.7428      & 0.0034     & -7.7552    & -7.7392     \\
Random      &           & -7.7611      & 0.0435     & -7.8428    & -7.5963     \\
Near-random &           & -7.7670      & 0.0010     & -7.7684    & -7.7636     \\
Near-zero   & \textsc{H$_2$O-8}     & -71.4633     & 0.0388     & -71.3888   & -71.5089    \\
Random      &           & -72.5497     & 0.7678     & -71.3903   & -73.2939    \\
Near-random &           & -72.2187     & 0.0000     & -72.2187   & -72.2187    
\end{tblr}
\end{table}

\section{Energy Evaluation}
\label{appendix:energy}
In this work, the energy at step $t$ is evaluated as the expectation value of the molecular Hamiltonian with respect to the quantum state prepared by the parameterized circuit, as defined in Eq.~\ref{eq:1} of the main text.

\subsection{Hamiltonian Representation}
The molecular Hamiltonian $H$ is first expressed in the second-quantized form using fermionic creation and annihilation operators. To simulate it on qubits, we map the fermionic operators to qubit operators through a fermion-to-qubit encoding. Two widely used schemes in literature are:
\begin{itemize}
    \item \textbf{Jordan--Wigner (JW) encoding}~(\cite{nielsen2010quantum, ostaszewski2021reinforcement}): maps each fermionic mode to a single qubit. Fermionic creation/annihilation operators are represented as strings of Pauli $Z$ operators followed by a local Pauli $X$ or $Y$, which ensures the correct anti-commutation relations.
    \item \textbf{Parity encoding}~(\cite{richardson2002efficient, wu2023quantumdarts}): encodes fermionic operators into qubits using cumulative parity information of occupation numbers, such that the $k$-th qubit stores the parity of the first $k$ modes. Compared to JW, this mapping often shortens the Pauli strings for operators acting on higher-index orbitals, and is particularly advantageous when combined with qubit tapering techniques that exploit system symmetries.
\end{itemize}

After such a transformation, the Hamiltonian takes the form:
\begin{equation}
H = \sum_{i} h_i P_i, \quad P_i \in \{I, X, Y, Z\}^{\otimes n},
\end{equation}
where each $P_i$ is an $n$-qubit Pauli string (a tensor product of single-qubit Pauli matrices $I, X, Y, Z$) and $h_i \in \mathbb{R}$ are coefficients determined by the molecular system and basis set.

\subsection{Expectation Value Computation}
At each step, the parameterized quantum circuit prepares a variational state $\ket {\psi(\vec{\theta})}$. Expanding Eq.~\ref{eq:1} with the Pauli decomposition of $H$, the energy can be written as:
\begin{equation}
E_t = \sum_i h_i \langle \psi(\vec{\theta})| P_i | \psi(\vec{\theta}) \rangle .
\end{equation}

In simulation, each expectation value $\langle P_i \rangle$ can be computed exactly from the state vector. On quantum hardware, these expectation values are estimated statistically: the circuit is executed multiple times, measurement outcomes in the relevant basis are collected, and the empirical averages are used to approximate $\langle P_i \rangle$. The weighted sum over all Pauli terms then gives the total energy estimate.
\section{Reinforcement Learning Environment Design} 
\label{appendix:components}

\subsection{Random Halting of the RL Environment}
\label{appendix:RH}

Following~\cite{patel2024curriculum}, the maximum number of steps per episode, $L$, defines the upper bound on the circuit construction process. Under the random halting (RH) scheme, a stochastic cap $\ell \le L$ is drawn from a negative binomial distribution~\cite{dougherty1990probability} parameterized by $L$ and a failure probability $p$, which specifies the maximum allowed length for the current episode. This stochastic cap introduces variability across episodes. Furthermore, if the agent constructs a circuit whose estimated energy falls below a given threshold, the episode may terminate early, resulting in an actual length shorter than $\ell$. The main motivation for RH is to encourage the agent to adapt to variable-length episodes and to improve exploration efficiency by enabling the discovery of more compact circuits in early successful episodes.

\subsection{Feedback-driven Curriculum Learning}
\label{appendix:FDCL}
Following~\cite{ostaszewski2021reinforcement}, we adopt a moving-threshold curriculum that adjusts a tolerance around an unattainable lower bound of the energy. Let $E_{\min}$ denote a proxy lower bound on the ground-state energy, which cannot be attained in practice. The threshold used in the main body is
\begin{equation}
\xi \;:=\; E_{\min} + \tau,
\end{equation}
where $\tau>0$ is a dynamic tolerance. Hence the reward condition $E_t<\xi$ is equivalent to $E_t - E_{\min} < \tau$.

\paragraph{Initialization.}
We initialize $\tau \leftarrow \xi_1$ (a hyperparameter), yielding $\xi = E_{\min}+\xi_1$. 
We also maintain the \emph{lowest energy} observed so far as a global quantity, initialized to the initial threshold level:
\begin{equation}
E_{\text{low}} \leftarrow \xi \quad\text{and updated at the end of every step by}\quad 
E_{\text{low}} \leftarrow \min\{E_{\text{low}},\, E_t\}.
\end{equation}

\paragraph{Greedy shift (periodic recalibration).}
Every $G$ episodes ($G$ is a hyperparameter), the tolerance is recalibrated using the current gap to the lower bound plus a small safety margin:
\begin{equation}
\tau \;\leftarrow\; \bigl|E_{\text{low}} - E_{\min}\bigr| + \delta,
\qquad\Rightarrow\qquad
\xi \;\leftarrow\; E_{\min} + \tau ,
\end{equation}
where $\delta>0$ provides slack. This periodic update typically \emph{tightens} the threshold by pushing it closer to the best energy found so far; however, if amortization has previously over-tightened the threshold, this recalibration can \emph{relax} it to a more achievable level.

\paragraph{Amortization (gradual tightening).}
When the agent attains a specified number of successful episodes (success count exceeding a threshold), the tolerance is gradually tightened:
\begin{equation}
\tau \;\leftarrow\; \tau - \frac{\delta}{\kappa},
\qquad\Rightarrow\qquad
\xi \;\leftarrow\; E_{\min} + \tau ,
\end{equation}
where $\kappa$ controls the number/size of incremental reductions. Counters are reset whenever a greedy shift occurs. This step raises task difficulty smoothly, encouraging finer improvements beyond the current $E_{\text{low}}$.

\paragraph{Summary.}
In practice, $E_{\text{low}}$ is tracked globally and updated after every step, while threshold updates (both greedy shifts and amortization) occur only at the end of each episode. Greedy shifts provide periodic downward \emph{recalibration} toward the current lowest energy, with a safety margin that can also \emph{relax} the threshold if it was over-tightened. Amortization then \emph{gradually tightens} the threshold between greedy shifts to promote steady refinement. Together, these mechanisms balance exploration depth and training stability without requiring prior knowledge of the true ground-state energy.

\subsection{Tensor-based Binary Circuit Encoding}
\label{appendix:TBCE}
\begin{figure}[ht]
    \centering
    \includegraphics[width=0.8\linewidth]{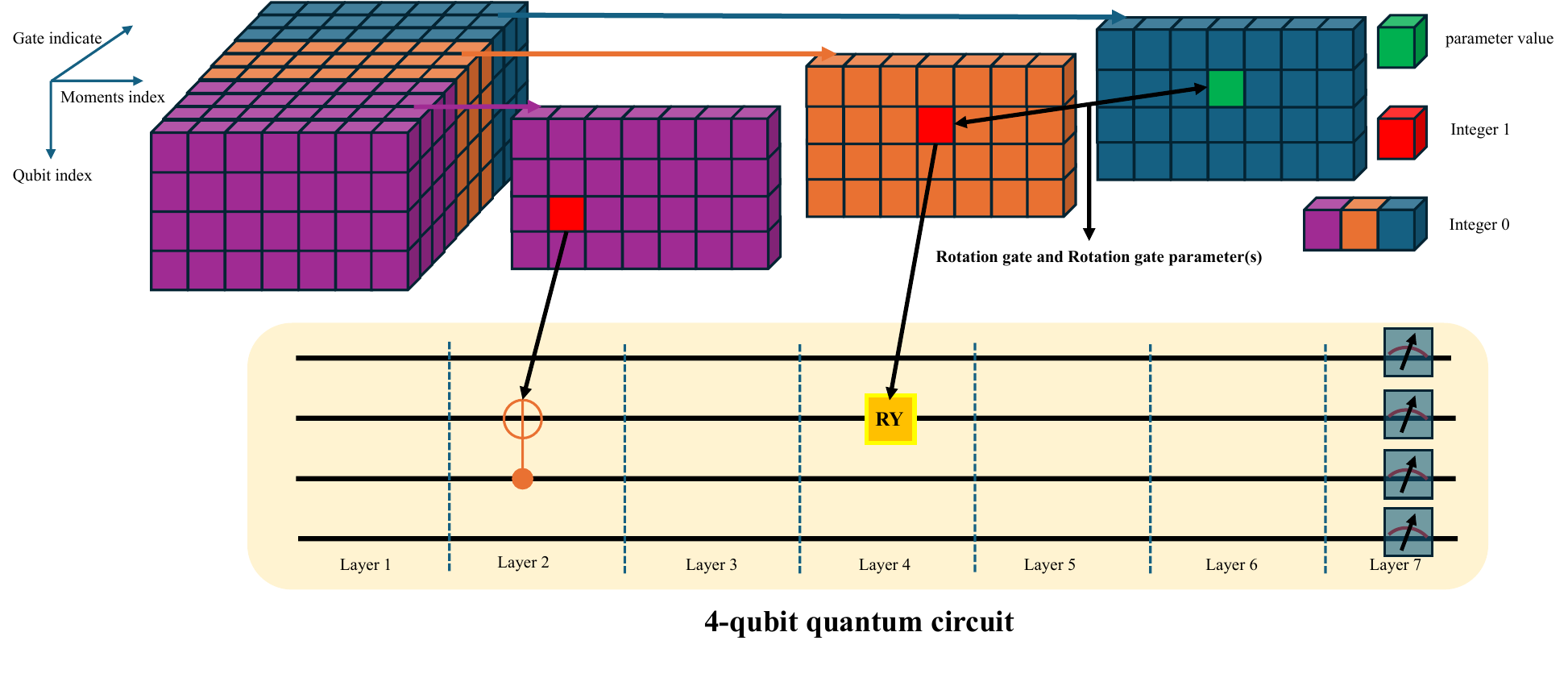}
    \caption{Illustration of the Tensor-based Binary Circuit Encoding for a 4-qubit quantum circuit. The purple and orange tensors store the position and type information of CNOT gates and rotation gates, respectively, where the \textit{Gate indicate} axis spans four dimensions for CNOTs and three dimensions for rotation gates. The red tensors further encode the parameters of the rotation gates. The \textit{Moments index} corresponds to the layer index in the quantum circuit, allowing a structured representation of gate placement and parameterization across different circuit layers.}
    \label{fig:example}
\end{figure}

Following~\cite{patel2024curriculum}, we represent the RL state using a tensorized encoding that captures the incremental construction of a quantum circuit.
While the original formulation employs a binary tensor of size $N \times (N+3) \times L$ to encode circuit structure, the proposed HyRLQAS framework extends this representation to incorporate gate parameter information, resulting in a state tensor of size $N \times (N+6) \times L$.

The last dimension indexes action slots, referred to as \emph{moments}. Rather than representing a fixed physical circuit depth, a moment is determined by a local stacking mechanism that specifies the earliest layer at which a newly selected gate can be placed without violating causal ordering along each qubit line.
This design allows gates acting on disjoint qubits to be assigned the same moment and thus executed in parallel.

At moment $\ell$, the leading $N \times N$ block encodes CNOT placement as a binary adjacency map, where entry $(i,j)=1$ indicates a CNOT with control qubit $i$ and target qubit $j$.
The subsequent $N \times 3$ block is a binary one--hot indicator for single--qubit rotations, with the three columns corresponding to $\mathrm{R}_X$, $\mathrm{R}_Y$, and $\mathrm{R}_Z$ gates.
The remaining $N \times 3$ block stores the associated rotation angles, with zero entries indicating the absence of a rotation gate.

Circuit layering follows a \emph{local stacking} rule.
When placing a single--qubit rotation on qubit $q$, the gate is assigned to the earliest available moment on that qubit.
When placing a CNOT on qubits $(i,j)$, the gate is assigned to the earliest moment that is valid for both qubits.
This rule enforces causal consistency while enabling parallel placement of gates on disjoint qubits.
Consequently, the realized circuit depth—defined by the maximum occupied moment—can be smaller than the number of gate placement decisions.
The fixed cap $L$ serves as a uniform upper bound on the episode horizon and ensures a consistent state dimension across episodes.

\subsection{Illegal Actions}
\label{appendix:IA}

Following~\cite{patel2024curriculum}, to prevent redundant or degenerate operations during circuit construction, the environment continuously maintains an \emph{illegal action list} throughout the entire building process. The list is initialized as empty and dynamically updated after each action executed by the agent. At every step, it is regenerated based on the current circuit configuration—specifically, the last gate applied on each qubit wire. Two main types of illegal actions are identified:

\paragraph{(1) Consecutive identical rotation gates.}
Applying two identical single-qubit rotation gates (e.g., $R_X$, $R_Y$, or $R_Z$) consecutively on the same qubit collapses to a single rotation with a combined angle:
\[
R_\alpha(\theta_1)R_\alpha(\theta_2) = R_\alpha(\theta_1 + \theta_2), \quad \alpha \in \{X, Y, Z\}.
\]
This operation introduces no new transformation and only increases circuit depth. 
Hence, if the last operation on a qubit is $R_\alpha$, the same rotation gate type on that qubit is added to the illegal action list for the next step.

\paragraph{(2) Inverted CNOT pairs.}
For two-qubit operations, a $\mathrm{CNOT}(q_c, q_t)$ gate immediately followed by its inverse $\mathrm{CNOT}(q_t, q_c)$ cancels out the entanglement effect and provides no additional expressive power. Accordingly, when a $\mathrm{CNOT}(q_c, q_t)$ is placed, the reversed pair $\mathrm{CNOT}(q_t, q_c)$ is marked as illegal until another valid gate is applied on either qubit.

\paragraph{Implementation.}
At each decision step, the agent receives from the environment an updated list of illegal actions determined by the current circuit configuration.
This list is converted into a binary mask $m_t^{\mathrm{IA}} \in \{0,1\}^{|\mathcal{A}_{\mathrm{disc}}|}$ over the discrete action space, where $m_{t,j}^{\mathrm{IA}} = 0$ indicates that action $j$ is invalid at step $t$, and $m_{t,j}^{\mathrm{IA}} = 1$ otherwise. Prior to sampling, the policy network suppresses all illegal actions by masking their corresponding logits, ensuring that the agent can only select valid gate placements from $\mathcal{A}_{\mathrm{disc}}$.

Once a valid action is executed, the illegal action list—and thus the mask $m_t^{\mathrm{IA}}$—is immediately updated according to the newly placed gate and the resulting circuit state. This dynamic and persistent filtering mechanism prevents redundant or semantically invalid gate sequences, enforces meaningful circuit growth, and significantly improves sample efficiency during training.

\section{Hybrid Action Space Implementation Details}
\label{HASID}
\subsection{Network and Discrete Action Space}
At each step $t$, the policy is parameterized by a neural network that first encodes the current state $s_t$ into a shared latent representation
\[
\mathbf{h}_t = \phi(s_t),
\]
where $\phi(\cdot)$ denotes a shared backbone network. This shared representation is then fed into two output heads: a discrete head that produces logits over gate candidates, and a continuous head that parameterizes Gaussian distributions for rotation parameters.
\begin{equation}
\boldsymbol{\ell}_t = \phi_{\mathrm{disc}}(\mathbf{h}_t), \qquad
(\boldsymbol{\mu}_t, \boldsymbol{\sigma}_t) = \phi_{\mathrm{cont}}(\mathbf{h}_t),
\end{equation}
where $\boldsymbol{\ell}_t \in \mathbb{R}^{|z_t |}$ and $\boldsymbol{\mu}_t, \boldsymbol{\sigma}_t \in \mathbb{R}^{|z_t|}$.

Before sampling the discrete action $z_t$, an illegal-action mask provided by the environment (Appendix~\ref{appendix:IA}) is applied to exclude invalid gate placements.
Specifically, the logits of illegal actions are clamped to a large negative constant $-C$ (approximating $-\infty$):
\[
\tilde{\ell}_{t,j} =
\begin{cases}
\ell_{t,j}, & \text{if action $j$ is legal at step $t$},\\
-C, & \text{otherwise},
\end{cases}
\]
so that illegal actions are effectively excluded from the sampling distribution. The discrete action is then sampled as
\[
z_t \sim
\mathrm{Categorical}\!\bigl(\mathrm{softmax}(\tilde{\boldsymbol{\ell}}_t)\bigr).
\]

For an $N$-qubit system, the discrete action space has cardinality $|z_t| = 2{N \choose 2} + 3N$, covering all ordered CNOT pairs and single-qubit rotations $\{R_X, R_Y, R_Z\}$ acting on each qubit.

If the selected discrete action $z_t$ corresponds to a parameterized rotation gate, the policy samples a full vector of candidate parameters from the Gaussian distributions parameterized by $(\boldsymbol{\mu}_t, \boldsymbol{\sigma}_t)$, and extracts the entry associated with the chosen discrete action as the initialization parameter:
\[
x_t \sim
\mathcal{N}\!\bigl(
\boldsymbol{\mu}_t[z_t],
\boldsymbol{\sigma}_t^{2}[z_t]
\bigr).
\]
For non-parameterized gates (i.e., CNOT), no continuous initialization is generated.

\subsection{Second Continuous Variable: Parameter Refinement}

Beyond initializing newly inserted rotation gates, HyRLQAS further supports \emph{parameter refinement} for previously placed gates. When a new gate is added, the parameters of existing rotation gates are adjusted to preserve optimization consistency as the circuit grows. This refinement is formulated as a conditional and factorized continuous component.

At step $t$, a vector of refinement increments is generated conditioned on the current circuit state $s_t$, the newly selected discrete gate $z_t$, and the newly parameter ${x}_t$. Formally, the refinement increment is sampled as
\[
\Delta\boldsymbol{x}_t \sim
\pi_{\mathrm{ref}}\!\left(
\cdot \mid s_t,\; z_t,\; {x}_t
\right),
\]
where $\Delta\boldsymbol{x}_t \in \mathbb{R}^{L}$ denotes a vector of additive parameter updates, with one entry corresponding to each prior construction step.

Each entry $\Delta x_t^{(i)}$ aligns with the rotation parameter introduced at
step $i$, following the temporal order of circuit construction.
To ensure semantic consistency, a binary \emph{delta mask}
$\delta_t \in \{0,1\}^L$ is maintained, where $\delta_t^{(i)} = 1$ if step $i$
placed a parameterized rotation gate and $\delta_t^{(i)} = 0$ otherwise.
The sampled refinement vector is filtered element-wise as
\[
\Delta\boldsymbol{x}_t \leftarrow \delta_t \odot \Delta\boldsymbol{x}_t,
\]
so that only valid rotation parameters are updated.

Refinement is applied additively:
\[
\boldsymbol{x}_{t+1} = \boldsymbol{x}_t + \eta \Delta\boldsymbol{x}_t,
\]
where $\eta$ is a step-size parameter. This additive refinement allows earlier parameters to be incrementally adjusted in response to newly introduced gates, tightly coupling circuit growth with parameter evolution and enabling structured exploration in the continuous parameter space.





\subsection{Hybrid Policy Optimization}

REINFORCE \citep{williams1992simple} is a classical reinforcement learning algorithm that learns a policy directly by updating its parameters to maximize expected cumulative reward. It is derived from the Policy Gradient Theorem, which provides an analytical expression for the gradient of the expected return with respect to the policy parameters. The algorithm seeks a parameterized policy $\pi_\theta(a | s)$—a probability distribution over actions given a state—that maximizes the objective $J(\theta) = \mathbb{E}[G_t | \pi_\theta]$. Here, the return $G_t$ from time $t$ is defined as the total discounted future reward:

$$
G_t = \sum_{k=0}^{T-t-1} \gamma^k r_{t+k+1},
$$

where $\gamma \in [0, 1]$ is the discount factor and $T$ denotes the episode length in episodic settings.

The HyRLQAS framework utilizes REINFORCE for policy optimization. This method is applied to both the discrete policy $\pi_{\phi}^d(z \mid s)$ and the continuous policy $\pi_{\phi}^a(x \mid s)$. We define a stochastic policy $\pi_{\phi}(z,x \mid s)$ that maps a state $s$ to both the discrete action $z$ and the continuous parameter $x$ as follows:

$$
\pi_{\phi}(z,x \mid s) = \pi_{\phi}^d(z \mid s) \, \pi_{\phi}^a(x \mid s),
$$

Let $\pi_{\phi}^d$ and $\pi_{\phi}^a$ denote the policy with parameters ${\phi}^d$ and ${\phi}^a$, and $J({\phi}^d)$ and $J({\phi}^a)$ denote the expected finite-horizon return of the policy. The gradient of $J({\phi}^d)$ is

$$\nabla_{\phi^d} J({\phi}^d)
=
\mathbb{E}\!\left[
\sum_t G_t \nabla_{\phi^d} \log \pi_{\phi^d}(z_t \mid s_t)
\right].$$

The gradient of $J({\phi}^a)$ is

$$\nabla_{\phi^a} J({\phi}^a)
=
\mathbb{E}\!\left[
\sum_t G_t \nabla_{\phi^a} \log \pi_{\phi^a}(x_t \mid s_t)
\right].$$

Accordingly, the policy parameters are updated by stochastic gradient ascent. At each training iteration, the discrete and continuous policy parameters are updated as
\[
\phi^d \leftarrow \phi^d + \alpha \sum_t G_t \nabla_{\phi^d} \log \pi_{\phi^d}(z_t \mid s_t)
\]
\[
\phi^a \leftarrow \phi^a + \alpha \sum_t G_t \nabla_{\phi^a} \log \pi_{\phi^a}(x_t \mid s_t)
\]
where $\alpha$ denotes the learning rate. The pseudo-code for the policy learning is presented in algorithm \ref{alg:hyrlqas}.

\begin{algorithm}[t]
\caption{HyRLQAS: Hybrid-Action Reinforcement Learning for Quantum Architecture Search}
\label{alg:hyrlqas}
\begin{algorithmic}[1]
\Require Hamiltonian $H$, maximum step limit $L$, policy $\pi_\phi$, external optimizer $\mathcal{O}$
\Ensure Trained policy $\pi_\phi$

\For{each training episode}
    \State Initialize empty circuit $c_0$, parameter state $\boldsymbol{x}_0 = \emptyset$, and state $s_0$
    \For{$t = 0$ to $L-1$}
        \State Sample discrete gate action with illegal-action masking:
        $z_t \sim \pi_\phi^d(\cdot \mid s_t)$
        \If{$z_t$ is parameterized}
            \State Sample initialization parameter:
            $x_t \sim \pi_\phi^a(\cdot \mid s_t)$
        \Else
            \State $x_t \leftarrow \emptyset$
        \EndIf
        \State Generate refinement increments $\Delta\boldsymbol{x}_t$ conditioned on $(s_t, z_t, x_t)$
        \State Apply additive refinement:
        $\boldsymbol{x}_t \leftarrow \boldsymbol{x}_t + \eta\,\Delta\boldsymbol{x}_t$
        \State Extend circuit structure:
        $c_{t+1} \leftarrow c_t \cup z_t$
        \State Invoke external optimizer:
        $\boldsymbol{x}_{t+1}^\star \leftarrow \mathcal{O}(c_{t+1}, H, \boldsymbol{x}_t)$
        \State Evaluate energy:
        $E_{t+1} = \langle \psi(c_{t+1}, \boldsymbol{x}_{t+1}^\star) \mid H \mid \psi(c_{t+1}, \boldsymbol{x}_{t+1}^\star) \rangle$
        \State Compute reward $r_t$ from $E_{t+1}$
        \State Update internal state:
        $s_{t+1} \leftarrow (c_{t+1}, \boldsymbol{x}_t)$
        \If{termination condition is satisfied}
            \State \textbf{break}
        \EndIf
    \EndFor
    \State Update discrete policy and continuous policy parameters using REINFORCE:
    \[
    \phi^d \leftarrow \phi^d + \alpha \sum_t G_t \nabla_{\phi^d} \log \pi_{\phi^d}(z_t \mid s_t)
    \]
    \[
    \phi^a \leftarrow \phi^a + \alpha \sum_t G_t \nabla_{\phi^a} \log \pi_{\phi^a}(x_t \mid s_t)
    \]
\EndFor
\end{algorithmic}
\end{algorithm}



\section{Experimental setup}
\subsection{Experimental Settings.}
\label{appendix:Exp_setting}

All methods adopt the COBYLA optimizer with a budget of 1{,}000 function evaluations per step for continuous parameter optimization.
For each method, we conduct five independent training runs using different random seeds sampled from $\{11111, \ldots, 55555\}$.
For each run, we select the checkpoint that achieves the lowest energy on the training trajectory as the final model, following the same protocol for all learning-based baselines. In the evaluation phase, the selected policy is kept fixed and no further parameter updates are performed.
All other evaluation settings are identical to those used during training.

We consider learning problems on 4-, 6-, and 8-qubit molecular systems. HyRLQAS employs a shared training protocol across all baselines, including the same policy network size, the same upper bound on circuit length, and identical environment settings for each problem size. To accommodate increasing problem complexity, these settings are scaled with the number of qubits, as summarized in Table~\ref{tab:settings}.

HyRLQAS further introduces a refinement backbone that is invoked after each gate insertion to adjust the parameters of previously placed gates.
It operates purely in the continuous parameter space and does not affect discrete gate selection. Therefore, all methods operate under a comparable architectural search space in terms of gate selection capacity. 

All experiments are conducted on a high-performance computing node equipped with two Intel Xeon Platinum 8592+ CPUs (128 physical cores, 256 threads, 2.6~GHz base / 3.9~GHz turbo) and eight NVIDIA L40S GPUs (48~GB memory each), running Ubuntu~22.04 LTS with PyTorch~2.8.0 built with CUDA~12.9.

\begin{table}
\centering
\caption{Training configuration across problem sizes.}
\label{tab:settings}
\begin{tblr}{
  width = \linewidth,
  colspec = {Q[96]Q[129]Q[137]Q[206]Q[365]},
  cells = {c},
  hline{1} = {-}{},
  hline{1, 2, 5} = {-}{0.12em},
}
Qubits & Episodes & Max Gate & Policy Network & HyRLQAS Refine Backbone                \\
4      & 10000    & 40       & {[}1000 $\times$ 5]   & {[}1000 $\times$  3]            \\
6      & 20000    & 70       & {[}2000 $\times$ 5]   & {[}2000 $\times$  3]            \\
8      & 40000    & 150      & {[}5000 $\times$ 5]   & {[}5000 $\times$  3]            \\
\end{tblr}
\end{table}

\subsection{Molecular Configuration}
\label{appendix:MoleConf}
Table~\ref{tab:molecule_config} summarizes the molecular geometries, fermion-to-qubit mappings, and symmetry reductions used in all experiments. All molecular systems are represented in the STO-3G basis, and all geometries are specified in Angstrom (\AA). Symmetry reduction (denoted as ``taper'') indicates whether qubit tapering based on $\mathbb{Z}_2$ symmetries is applied.

\begin{table}
\centering
\small
\caption{Molecular geometries, fermion-to-qubit mappings, and symmetry reductions used in the experiments.}
\label{tab:molecule_config}
\begin{tblr}{
  cells = {c},
  hline{1-2,7} = {-}{0.12em},
}
Molecule  & Geometry                                                       & Mapping        & taper \\
\textsc{H$_2$-4}   & H (0, 0, 0.35);H (0, 0, -0.35)                                 & Jordan-Wigner  & Yes   \\
\textsc{LiH-4}     & Li (0, 0, 0); H (0, 0, 3.4)                                    & parity         & Yes   \\
\textsc{LiH-6}     & Li (0, 0, 0); H (0, 0, 2.2)                                    & jordan\_wigner & No    \\
\textsc{BEH$_2$-6} & Be (0, 0, 0); H( 0, 0, 1.33); H( 0, 0, -1.33)                  & jordan\_wigner & No    \\
\textsc{H$_2$O-8}  & H(-0.021, -0.002, 0); O(0.835, 0.452, 0); H (1.477, -0.273, 0) & jordan\_wigner & No      
\end{tblr}
\end{table}

\section{Experimental under nosiy}
\label{appendix:Exper_noisy}
\subsection{Noisy simulation settings}
To assess robustness under realistic noise, we conduct supplementary experiments using depolarizing noise models. Specifically, single-qubit and two-qubit depolarizing channels are applied after each corresponding gate operation. We consider two noise regimes: a \emph{realistic} setting with single-qubit depolarizing probability $p_1=0.001$ and two-qubit depolarizing probability $p_2=0.01$, and an \emph{aggressive} setting with $p_1=0.01$ and $p_2=0.05$, which are consistent with prior noisy VQE and QAS studies.

Due to computational constraints, noisy simulations are conducted only for HyRLQAS and two representative RL-based QAS baselines (CRLQAS and VanillaRL). All methods are trained under the same noise setting, where depolarizing noise is injected during circuit execution to compute the reward. For evaluation, we follow prior work~\citep{patel2024curriculum} and report results obtained from noiseless measurements of the learned circuits, so as to assess architectural quality independently of stochastic noise effects.

\subsection{Results under noise}
Table~\ref{Tab:Main_Exp_noise} reports the energy error, circuit depth, and gate counts obtained for \textsc{LiH-4} and \textsc{H$_2$-4} under two depolarizing noise regimes.
Across both realistic and aggressive noise settings, HyRLQAS consistently matches or outperforms the RL-based baselines in terms of energy error, while producing circuits with reduced depth and fewer two-qubit gates. For \textsc{LiH-4}, HyRLQAS achieves the lowest energy error among all methods under both noise levels, improving upon CRLQAS and VanillaRL by up to a factor of $2\!\sim\!3$. For the simpler \textsc{H$_2$-4} system, all methods reach near-identical energy accuracy, whereas HyRLQAS consistently yields more compact circuits with fewer entangling operations. These supplementary results suggest that the advantages of HyRLQAS observed in noiseless settings are preserved under the moderate noise levels considered in this study.

\begin{table*}[ht]
\centering
\scriptsize
\caption{Comparison of HyRLQAS with baseline methods under noisy quantum simulations. Results are shown for \textsc{LiH-4} and \textsc{H2-4} under realistic and aggressive noise models, reporting energy error, circuit depth, and gate counts}
\label{Tab:Main_Exp_noise}
\resizebox{0.9\textwidth}{!}{
\begin{tblr}{
  cells = {c},
  cell{1}{1} = {r=2}{},
  cell{1}{3} = {c=4}{},
  cell{1}{7} = {c=4}{},
  cell{3}{1} = {r=3}{},
  cell{6}{1} = {r=3}{},
  hline{1,3,9} = {-}{0.12em},
  hline{2} = {2-11}{0.12em},
  hline{6} = {1-11}{0.12em},
  vline{1,2,3,7,11} = {-}{0.12em},
}
Molecule & Nosie level  & Realistic(1-qubit depol = 0.001 and 2-qubit depol =0.01) &       &      &     & Aggressive(1-qubit depol = 0.01 and 2-qubit depol =0.05) &       &      &     \\
         & Method       & Error                                                    & Depth & CNOT & ROT & Error                                                    & Depth & CNOT & ROT \\
\textsc{LiH-4}    & Ours HyRLQAS & 1.1×10\^{}-3                                             & 26    & 12   & 15  & 5.2×10\^{}-3                                             & 10    & 3    & 14  \\
         & CRLQAS       & 3.8×10\^{}-3                                             & 16    & 11   & 21  & 9.4×10\^{}-3                                             & 14    & 13   & 17  \\
         & Vanilla RL   & 2.9×10\^{}-3                                             & 24    & 12   & 23  & 6.1×10\^{}-3                                             & 17    & 19   & 10  \\
\textsc{H2-4}     & Ours HyRLQAS & 1.31×10\^{}-8                                            & 4     & 3    & 4   & 1.34×10\^{}-8                                            & 5     & 6    & 5   \\
         & CRLQAS       & 1.38×10ˆ-8                                               & 14    & 12   & 8   & 1.39×10\^{}-8                                            & 11    & 9    & 13  \\
         & Vanilla RL   & 1.34×10ˆ-8                                               & 17    & 6    & 10  & 3.15×10\^{}-8                                            & 11    & 11   & 10       
\end{tblr}
}
\end{table*}

\section{Refinement Dynamics During Circuit Construction}
\label{app:refine_magnitude}

\begin{figure}[t]
    \centering
    \begin{subfigure}{0.48\linewidth}
        \centering
        \includegraphics[width=\linewidth]{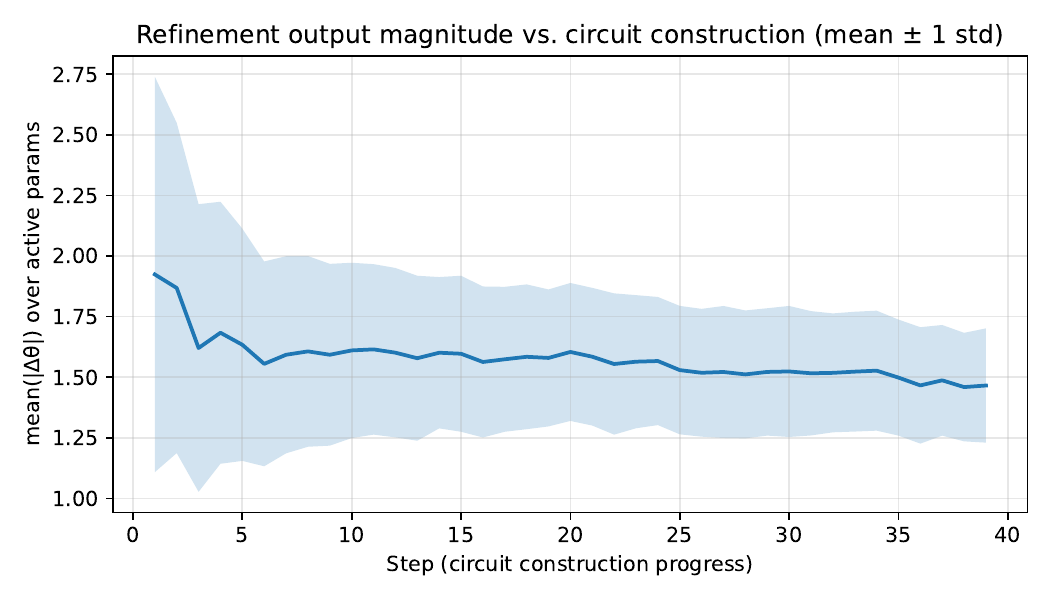}
        \caption{\textsc{LiH-4} (construction step)}
        \label{fig:refine_step_lih4}
    \end{subfigure}
    \hfill
    \begin{subfigure}{0.48\linewidth}
        \centering
        \includegraphics[width=\linewidth]{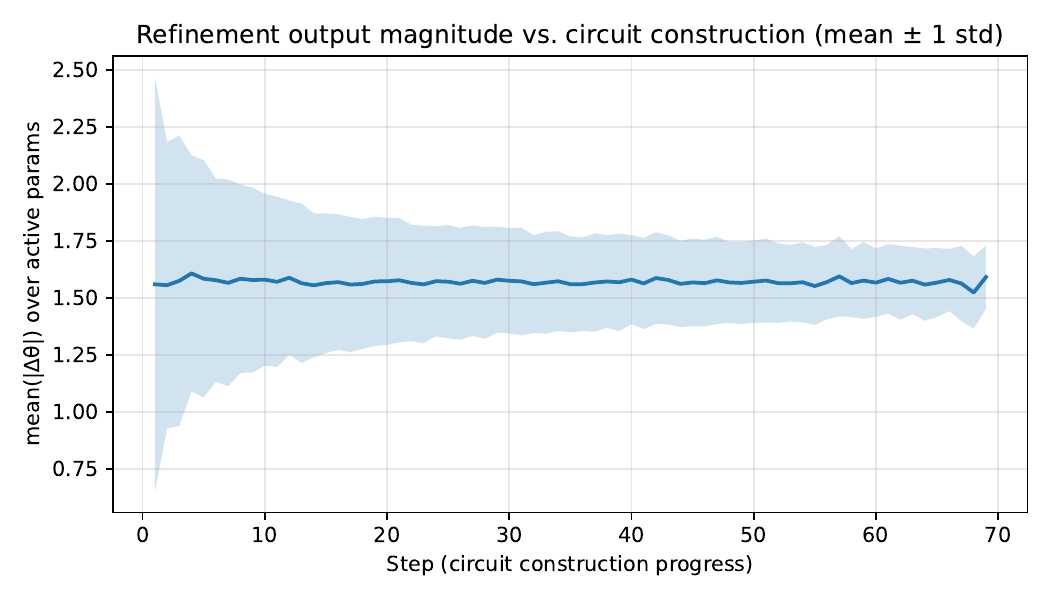}
        \caption{\textsc{LiH-6} (construction step)}
        \label{fig:refine_step_lih6}
    \end{subfigure}

    \vspace{0.6em}

    \begin{subfigure}{0.48\linewidth}
        \centering
        \includegraphics[width=\linewidth]{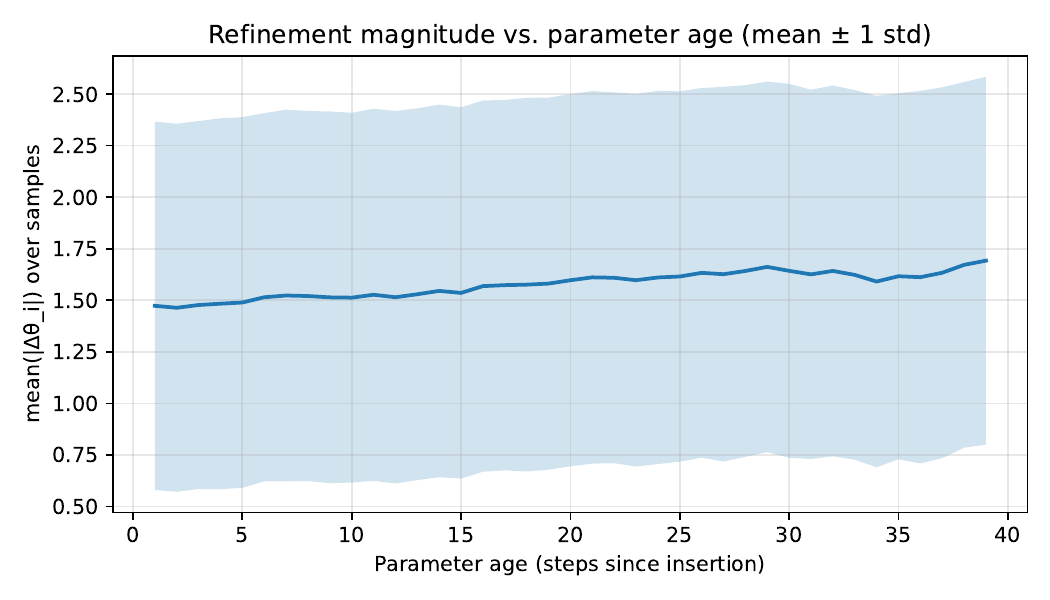}
        \caption{\textsc{LiH-4} (parameter age)}
        \label{fig:refine_age_lih4}
    \end{subfigure}
    \hfill
    \begin{subfigure}{0.48\linewidth}
        \centering
        \includegraphics[width=\linewidth]{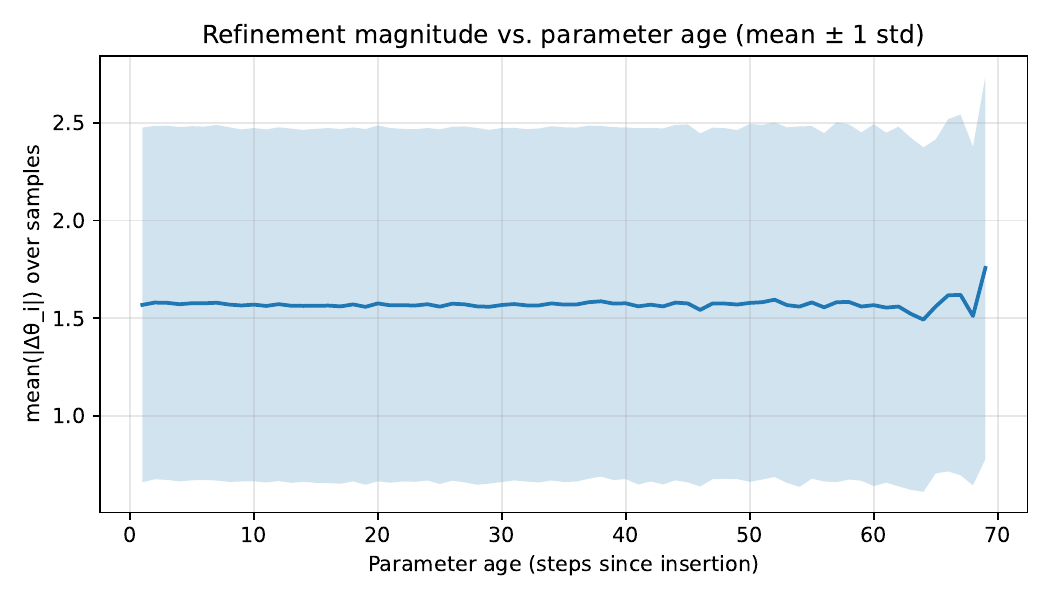}
        \caption{\textsc{LiH-6} (parameter age)}
        \label{fig:refine_age_lih6}
    \end{subfigure}

    \caption{
    Refinement magnitude analysis from two complementary perspectives.
    \textbf{Top row:} mean absolute refinement magnitude per active rotation parameter as a function of circuit construction step.
    \textbf{Bottom row:} mean absolute refinement magnitude as a function of parameter age (number of steps since insertion).
    }
    \label{fig:refine_combined}
\end{figure}

Figure~\ref{fig:refine_combined} characterizes the behavior of the refinement mechanism from two complementary perspectives:
(i) its evolution over the circuit construction process (top row),
and (ii) its effect on parameters of different lifetimes (bottom row).

All results are obtained by evaluating a \emph{fixed, trained policy checkpoint} in test mode over \textbf{1000 independent episodes}, where each episode corresponds to a complete circuit construction trajectory. At each construction step $t$, the refinement module produces additive updates $\Delta\boldsymbol{x}_t$ for all currently active rotation parameters.

\paragraph{Refinement behavior over construction steps.}
The top row of Figure~\ref{fig:refine_combined} reports the mean absolute refinement magnitude per active rotation parameter as a function of circuit construction progress. To decouple refinement behavior from the increasing parameter dimensionality as the circuit grows, we normalize by the number of active rotation parameters at each step. Across both molecular systems, refinement exhibits a short transient phase followed by rapid
stabilization. Despite increasing circuit depth and parameter count, the magnitude of refinement updates remains bounded and does not grow with construction length, indicating that refinement does not introduce step-dependent instability.

\paragraph{Refinement behavior over parameter lifetime.}
The bottom row of Figure~\ref{fig:refine_combined} groups refinement updates by parameter age, defined as the number of construction steps elapsed since a rotation parameter was introduced.

For \textsc{LiH-6}, the refinement magnitude remains approximately constant across parameter ages, indicating that parameters introduced earlier do not receive increasingly stronger updates as the circuit grows. In contrast, \textsc{LiH-4} exhibits a gradual increase in refinement magnitude with parameter age. Importantly, this increase is smooth and bounded, with no evidence of divergence or instability. This suggests that, in smaller systems, refinement continues to adapt earlier parameters to accommodate later circuit growth, while maintaining controlled update scales.

Overall, both systems demonstrate that refinement avoids uncontrolled amplification with parameter age, supporting stable and coherent parameter evolution during incremental circuit construction.

\section{Appendix: Visualization of Final Optimized Architectures}
\label{appendix:final_arch}
\begin{figure*}[]
    \centering
    \includegraphics[width=0.90\linewidth]{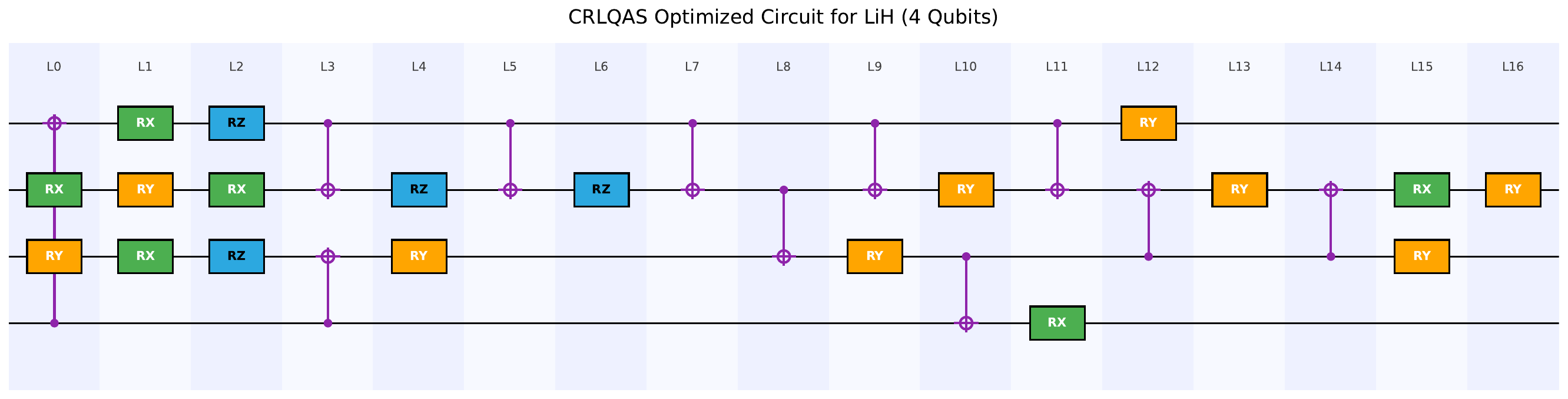}
    \vspace{0.8em}
    \includegraphics[width=0.90\linewidth]{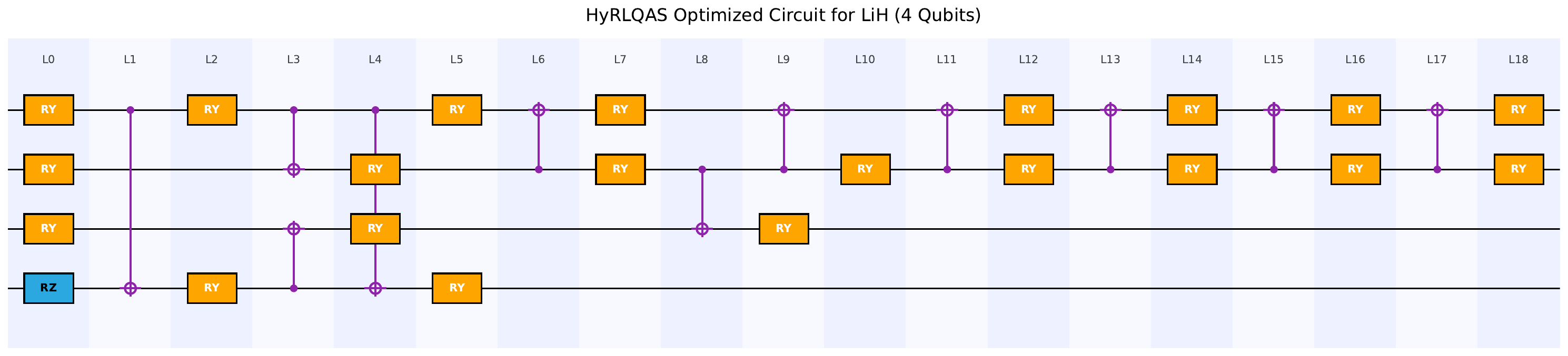}
    \vspace{0.8em}
    \includegraphics[width=0.90\linewidth]{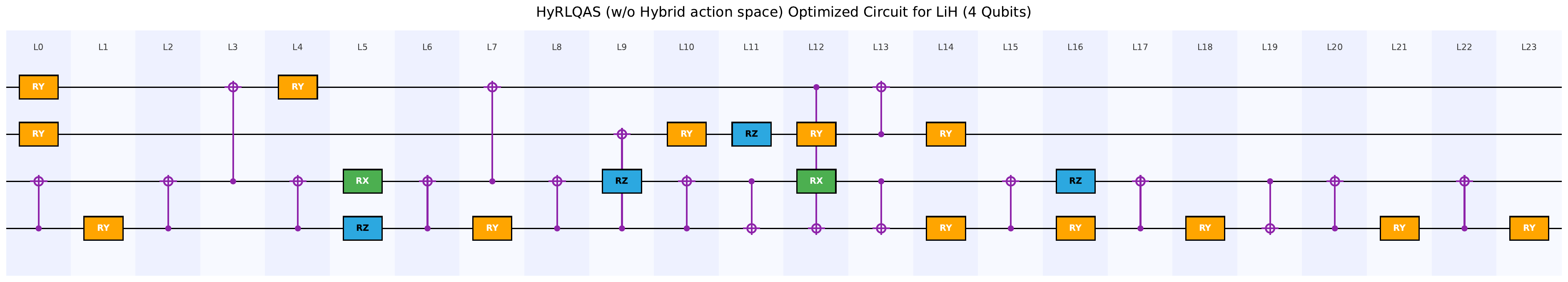}
    \vspace{0.8em}
    \includegraphics[width=0.90\linewidth]{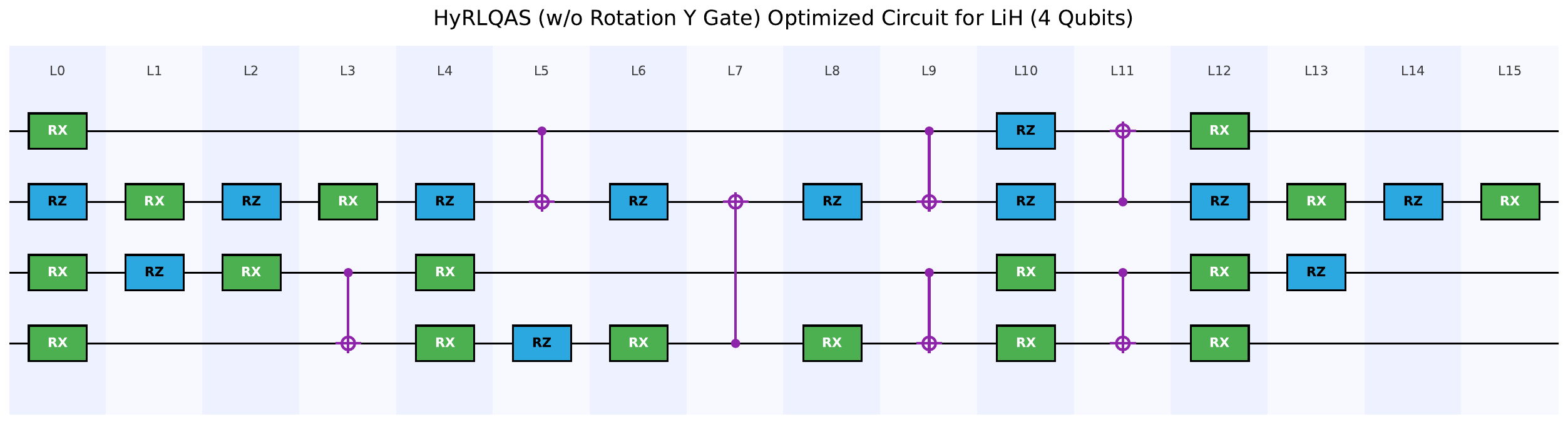}
    \caption{Comparison of optimized circuit architectures for the \textsc{LiH} (4-qubit) benchmark.
    Top: CRLQAS. 
    Second: full HyRLQAS.
    Third: HyRLQAS without the hybrid action space (A1).
    Bottom: HyRLQAS with $R_Y$ rotations masked.}
    \label{fig:architecture_comparison}
\end{figure*}
\subsection{Our HyRLQAS Vs CRLQAS}
We compare the circuit architectures produced by CRLQAS and HyRLQAS for the same molecular benchmark based on the circuit visualizations. The key differences are structural rather than purely resource-based.

\paragraph{Gate-set composition at fixed visible rotation budget.}
From the plotted circuits, both methods allocate the same number of single-qubit rotation blocks (19 total). However, the gate \emph{type} distribution differs sharply: CRLQAS employs a heterogeneous mixture of $R_X$, $R_Y$, and $R_Z$ rotations ($N_{R_X}=6$, $N_{R_Y}=9$, $N_{R_Z}=4$), whereas HyRLQAS is overwhelmingly dominated by $R_Y$ ($N_{R_X}=1$, $N_{R_Y}=17$, $N_{R_Z}=1$).
Thus, HyRLQAS does not simply increase the apparent number of variational degrees of freedom; instead, it concentrates the local parameterization into a restricted (predominantly single-axis) family.

\paragraph{Layer motif and relation to real-amplitude hardware-efficient templates.}
The HyRLQAS architecture visually resembles a hardware-efficient motif of alternating local rotation layers and entangling layers, with local updates dominated by $R_Y$.
This structure is closely related in spirit to the commonly used \texttt{RealAmplitudes} template, which consists of alternating layers of $R_Y$ rotations and CX entanglers and is explicitly designed to prepare states with purely real amplitudes \cite{qiskit_realamplitudes}.
Although HyRLQAS includes a small number of non-$R_Y$ rotations in the plotted circuit (one $R_X$ and one $R_Z$), the strong bias toward $R_Y$ suggests that the search favours a near-real-amplitude parameterization.

\paragraph{Symmetry-aligned inductive bias (conditional statement).}
For time-reversal invariant Hamiltonians represented in an appropriate real basis, eigenfunctions can be chosen real in standard treatments \cite{mit_time_reversal_notes}.
Under these conditions, exploring phase degrees of freedom introduced by frequent $R_Z$ rotations may be unnecessary for accurately representing the ground state.
Consequently, restricting the ansatz toward a real(-dominant) submanifold can reduce redundant directions in parameter space and improve practical optimization behaviour, without sacrificing the ability to approximate the physically relevant solution.

\paragraph{Expressivity--trainability trade-off and architecture search.}
Variational quantum algorithm performance is known to depend strongly on ansatz design, with trainability limitations (including barren plateau phenomena in broad settings) constituting a central concern \cite{mcclean2018barren,cerezo2021variational}.
Quantum architecture search (QAS) is explicitly motivated by the need to balance expressivity against trainability and (on hardware) noise accumulation; increasing circuit resources may increase expressivity while degrading trainability, and QAS aims to identify architectures that achieve good performance within these constraints \cite{du2022quantum}.
In this context, the HyRLQAS preference for a restricted local gate alphabet (dominantly $R_Y$) can be interpreted as a form of search-induced regularization: the learned circuit family may be ``just expressive enough'' for the task while being easier to optimize than a more heterogeneous alternative.

\paragraph{Recommended quantitative substantiation.}
To strengthen the structural interpretation, one may quantify (i) expressibility and entangling capability using established descriptors \cite{sim2019expressibility}, and (ii) the entanglement generated by the optimized states via a compact multipartite entanglement summary such as the Meyer--Wallach measure \cite{meyer2001global}.
If the final circuits are classically simulable at the studied size, an additional direct test of the near-real-amplitude hypothesis is to evaluate a "realness" diagnostic of the optimized state (e.g., $\|\mathrm{Im}(\psi)\|_2$).

\subsection{Effect of masking $R_Y$ rotations}
\label{appendix:final_arch_NoRY}

To verify whether the strong $R_Y$ dominance observed in HyRLQAS architectures is functionally important rather than incidental, we conduct a small controlled experiment in which all $R_Y$ rotation actions are masked during training, while keeping the environment, optimization procedure, and circuit depth budget unchanged.

Under this constraint, the resulting circuit reaches a final energy error of $0.0031$, which is notably worse than the full HyRLQAS model.
The learned architecture is substantially more heterogeneous, consisting of $17$ $R_X$ gates, $12$ $R_Z$ gates, and $7$ CNOT gates, with an effective depth of $16$. The corresponding circuit visualization is included in Fig.~\ref{fig:architecture_comparison}.

Despite retaining comparable circuit depth, removing $R_Y$ rotations leads to clear performance degradation. 
This suggests that the $R_Y$-dominant structures discovered by HyRLQAS are not merely a byproduct of the action space, but play an active role in shaping an optimization-friendly ansatz.
In particular, masking $R_Y$ forces the agent to rely on rotation axes that introduce additional phase structure, plausibly pushing the search away from a near-real-amplitude regime that is empirically favorable for this problem.

\subsection{Effect of removing the hybrid action space (ablation A1).}
\label{appendix:final_arch_A1}
Comparing the optimized circuits produced by \textbf{HyRLQAS} and its ablation variant without the hybrid action space (A1)(Sec~\ref{sec:ablationA}), we observe a clear structural change in the resulting architectures. While the full \textbf{HyRLQAS} model learns a highly regular motif dominated by $R_Y$ rotations interleaved with a relatively consistent entangling pattern, the A1 variant produces a visibly more heterogeneous circuit structure, with increased use of $R_X$ and $R_Z$ rotations and a less regular placement of entangling gates. This suggests that removing the hybrid action space weakens the inductive bias toward structured, near-real-amplitude ansatz families, leading to architectures that are less constrained and potentially harder to optimize. The observed performance degradation in ablation A1 is therefore consistent with the corresponding loss of structural regularity, rather than being attributable to differences in overall circuit depth or gate count.

\section{Quantum Neural Tangent Kernel (QNTK)}
\label{app:qntk}

\subsection{Definition}

Consider a parametrized quantum circuit (PQC) $U(\boldsymbol{\theta})$ acting on an initial state $\ket{0}$, which produces the quantum state
\[
\ket{\psi(\boldsymbol{\theta})} = U(\boldsymbol{\theta}) \ket{0}.
\]
Let $O$ be a Hermitian observable. The model output is defined as
\[
f(\boldsymbol{\theta}) 
= \langle \psi(\boldsymbol{\theta}) | O | \psi(\boldsymbol{\theta}) \rangle.
\]

Following the Quantum Neural Tangent Kernel (QNTK) framework~\cite{liu2022representation}, the QNTK formally corresponds to a kernel matrix defined by gradient inner products of model outputs with respect to circuit parameters. 
In this work, we use a scalar proxy of the QNTK, defined as the squared norm of the gradient:
\[
K_1 (\boldsymbol{\theta}) 
= \| \nabla_{\boldsymbol{\theta}} f(\boldsymbol{\theta}) \|_2^2
= \sum_{i=1}^d 
\left(\frac{\partial f(\boldsymbol{\theta})}{\partial \theta_i}\right)^2,
\]
where $d$ is the number of trainable parameters.

\noindent
Note that the formal QNTK is a kernel matrix
$K_{ij} = \sum_{\ell} \partial_{\theta_\ell} z_i \partial_{\theta_\ell} z_j$,
where $z_i$ denotes model outputs on different inputs~\cite{liu2022representation}.
The quantity $K_1$ used here corresponds to the trace of this kernel and serves as a compact indicator of overall trainability.

This scalar measure reflects the sensitivity of the quantum output with respect to parameter perturbations and acts as a first-order local descriptor of optimization geometry.

\subsection{Second-Order QNTK}

To capture curvature effects beyond first-order sensitivity, we further define a second-order QNTK as the squared Frobenius norm of the Hessian matrix:
\[
K_2 (\boldsymbol{\theta}) = \left\| \nabla_{\boldsymbol{\theta}}^2 f(\boldsymbol{\theta}) \right\|_F^2.
\]

This quantity is closely related to the quantum meta-kernel (dQNTK) proposed in~\cite{liu2022representation}, which characterizes leading-order nonlinearity and representation learning effects beyond the frozen-kernel regime.
Large values of $K_2$ indicate rapidly varying gradients and potentially rugged optimization landscapes, whereas smaller or more concentrated $K_2$ correspond to locally smoother geometry.

\subsection{Experimental Details}

For each variational circuit generated by HyRLQAS, we evaluate QNTK statistics under four parameter configurations: 
policy-guided initialization, policy-guided initialization without refinement accumulation, zero initialization, and random initialization.

All QNTKs are computed immediately after circuit generation and before any classical parameter optimization.
This design isolates architectural and initialization effects from optimizer dynamics.

For each configuration, we compute both $K_1$ and $K_2$ at the generated parameter point.
We report the empirical distributions of $K_1$ and $K_2$, as well as pairwise differences $\Delta K_1$ and $\Delta K_2$ between matched configurations.
All results shown in Figure~4 are aggregated over the same set of circuits produced by the frozen policy, ensuring that differences arise solely from parameterization strategy.

\subsection{Interpretation}

The QNTK provides a local probe of optimization geometry.
Larger $K_1$ implies stronger descent directions and increased parameter sensitivity, whereas reduced or concentrated $K_2$ reflects improved numerical conditioning and smoother curvature.

In particular, a nearly constant or narrow $K_1$ distribution suggests a frozen (lazy-training) regime, while increased variation in $K_2$ indicates deviation from linearized dynamics and the onset of representation learning~\cite{liu2022representation}.
Therefore, shifts in QNTK statistics reveal how hybrid-action learning modifies the underlying landscape, providing insight into the improved convergence behavior observed in the main experiments.


\paragraph{Discussion.}
While the external optimizer accounts for the majority of the wall-clock time per step, its cost scales with problem difficulty rather than policy complexity.
The consistently low inference time demonstrates that the hybrid-action policy introduces negligible overhead.
Instead, the dominant runtime is governed by classical optimization, highlighting the importance of effective policy-guided initialization to reduce the number of required optimization iterations.


\end{document}